\documentclass[conference]{IEEEtran}
\IEEEoverridecommandlockouts
\usepackage[square,sort&compress,comma,numbers]{natbib}
\usepackage{amsmath,amssymb,amsfonts}
\usepackage{graphicx}
\usepackage{textcomp}
\usepackage{xcolor}
\usepackage{url}
\usepackage{algorithm}
\usepackage{multirow}
\usepackage{verbatim}
\usepackage{graphicx}
\usepackage{tabularx}
\usepackage{subfigure}
\usepackage{booktabs}
\usepackage{xspace}
\usepackage{enumerate}
\usepackage{shortvrb}
\usepackage[noend]{algpseudocode}
\usepackage{mathtools}
\usepackage{url}

\usepackage[all]{nowidow}
\usepackage{balance}
\newcommand{\eat}[1]{}

\usepackage{amsthm}
\theoremstyle{definition}
\newtheorem{definition}{Definition}
\newtheorem{lemma}{Lemma}
\def\BibTeX{{\rm B\kern-.05em{\sc i\kern-.025em b}\kern-.08em
    T\kern-.1667em\lower.7ex\hbox{E}\kern-.125emX}}

\newcommand{\metric}[1]{{\mbox{#1}}}
\newcommand{\Pat}[1]{\metric{P@{$#1$}}}
\newcommand{\var}[1]{\mbox{\em #1}}

\newcommand{\myurl}[1]{{\url{#1}}}

\newcommand{\myparagraph}[1]{\vspace{0.2\baselineskip}\noindent{\textit{#1}}.~}

\newcommand{\noi}{\noindent}
\newcommand{\mycomment}[1]{}

\newlength{\onedigit}
\newcounter{todocount}
\newcommand{\Oc}{producer}
\newcommand{\User}{consumer}

\newcommand{\Tree}{CPPse-index}
\newcommand{\up}{\mbox{$u^{p}$}}
\newcommand{\uc}{\mbox{$u^{c}$}}
\newcommand{\uci}[1]{\mbox{$u^{c}_{#1}$}}
\newcommand{\func}[1]{\textsf{\small{#1}}}
\newcommand{\dataset}[1]{\mbox{\textsf{\small{#1}}}}
\newcommand{\datasetcap}[1]{\mbox{\textsf{\scriptsize{#1}}}}
\newcommand{\movielen}{\dataset{MLens}}
\newcommand{\youtube}{\dataset{YTube}}
\newcommand{\smovielen}{\dataset{SynMLens}}
\newcommand{\syoutube}{\dataset{SynYTube}}
\newcommand{\movielenc}{\datasetcap{MLens}}
\newcommand{\youtubec}{\datasetcap{YTube}}
\newcommand{\smovielenc}{\datasetcap{SynMLens}}
\newcommand{\syoutubec}{\datasetcap{SynYTube}}

\setcitestyle{citesep={,},open={},close={}}

\makeatletter
\def\NAT@def@citea{\def\@citea{\NAT@separator}}
\makeatother
\begin{document}

\title{Online Social Media Recommendation over Streams\thanks{This paper appears at ICDE 2019}}
\author{
{Xiangmin Zhou{\small $~^{\dagger1}$}, Dong Qin{\small $~^{\dagger2}$},  Xiaolu Lu{\small $~^{\dagger3}$},

Lei Chen{\small $~^{*4}$}, Yanchun Zhang{\small $^{\ddagger5}$} }
\vspace{1.6mm}\\
\fontsize{10}{10}\selectfont\itshape
$^{\dagger}$\,RMIT University, Melbourne, Australia\\
\fontsize{9}{9}\selectfont\ttfamily\upshape
$^{123}$\,\{xiangmin.zhou,dong.qin,xiaolu.lu\}@rmit.edu.au
\vspace{1.2mm}\\
\fontsize{10}{10}\selectfont\rmfamily\itshape
$^{*}$\,Hong Kong University of Science and Technology,
Hong Kong, China\\
\fontsize{9}{9}\selectfont\ttfamily\upshape
$^{4}$\,leichen@cse.ust.hk
\vspace{1.2mm}\\
\fontsize{10}{10}\selectfont\rmfamily\itshape
$^\ddagger$\,Victoria University,
Melbourne, Australia\\
\fontsize{9}{9}\selectfont\ttfamily\upshape
$^{5}$\,yanchun.zhang@vu.edu.au
}

\maketitle

\begin{abstract}

As one of the most popular services over online communities, the social recommendation has attracted increasing research efforts recently. Among all the recommendation tasks, an important one is social item recommendation over high speed social media streams. Existing streaming recommendation techniques are not effective
for handling social users with diverse interests. Meanwhile, approaches for recommending items
to a particular user are not efficient when applied to a huge number of users over high
speed streams. In this paper, we propose a novel framework for the social recommendation over streaming
environments. Specifically, we first propose a novel Bi-Layer Hidden Markov Model (BiHMM) that
adaptively captures the behaviors of social users and their interactions with influential
official accounts to predict their long-term and short-term interests. Then, we design a new
probabilistic entity matching scheme for effectively identifying the relevance score of a streaming
item to a user. Following that, we propose a novel indexing scheme called {\Tree} for improving the
efficiency of our solution. Extensive experiments are conducted to prove the high performance of
our approach in terms of the recommendation quality and time cost.
\end{abstract}


\section{Introduction}\label{sec:recommendation}
With the explosive growth of online service platforms, an increasing number of people and
enterprises are undertaking personal and professional tasks online. Recent statistics shows there are now 15 million active Australians on Facebook, which is 60\% of the Australian population \cite{socialmedianews}. The digital universe is doubling in size every two years, and by 2020 the data users create and copy annually will reach 44 trillion gigabytes \cite{2014iview}. In order for organizations, governments, and individuals to understand their users, and promote their products or services, it is necessary for them to analyse these social data and recommend the media or online services in real time. A large volume of social media are proliferated in the form of streams, which has raised the demand of online media stream recommendation. Recommending streaming items over social communities is very important for many applications such as entertainment, online product promotion, and news broadcasting. For instance, the fans can enjoy their idols' performances once they are available online by continuously receiving the recommendations from the system over the dynamically changing social networks such as YouTube. An online company may accelerate the propagation of its digital commercials via the stream recommender systems to potential customers to boost the sales of their products. For news broadcasting, users can be notified in time what is happening moment by moment, and take prompt action in crises. Practically, these applications are time-critical, which demands the development of efficient stream recommendation approaches.

We study the problem of continuous recommendation over social communities. Given a new incoming social item $v$, a relevance function on social item and users, we aim to deliver the item $v$ to the top $k$ users that have the highest relevance scores. For example, a clip on a new KFC dessert can be broadcasted to the top interested users immediately after the uploading, which directly increases the product purchase and brand recall. For stream recommendation, three key issues need to be addressed. First, we need to construct a robust model that effectively predicts the short-term and long-term interests of different social users. While users' long-term interests keep relatively stable, their short-term interests can be changed rapidly due to the frequent social activities. Users' behaviors can be affected by their previous activities and their interacted media producers as well. For instance, a user interested in football games may become interested in music after watching a broadcasting from a producer on the family of David Beckham and Victoria Beckham. A good model should be able to capture the users' temporal involvement over their own activities and their media producers to reflect users' current preferences for high quality recommendation. Then, we need to design a novel solution for matching the streaming items with social users. As a large number of near duplicate items may appear in media streams, it is unreasonable to recommend them to a target user repeatedly. For example, John watched a video of Refael Nadal in Australian Open 2018. He may get bored after watching Nadal's videos repeatedly. Probably, he is interested in the videos on other tennis players as well, such as Roger Federer and Maria Sharapova. A good item-user matching approach should be able to recommend diverse items to an interested user. Finally, we need to design an efficient index scheme for searching the interested users with respect to an incoming item. According to the statistics from Hootsuite \cite{hootsuite2018}, YouTube has more than 1.5 billion users in 2018, and the number is increasing annually. Obviously, sequentially matching each incoming item with this huge number of users is infeasible for the efficient recommendation.

Based on the evaluation objectives of recommendation, the previous social recommendation approaches can be classified into two categories, relevance-based \cite{zhou2015online,DBLP:journals/vldb/ZhouCZQCHW17,balakrishnan2018using,li2018personalized,yang2017collaborative} and diversity-based \cite{ribeiro2015multiobjective,he2012gender,hurley2013personalised,castells2015novelty,zanitti2018user}. Relevance-based approaches identify the most similar items matched with a user predefined profile based on the present content and context features, producing a list of items relevant to the ones viewed by this user in the past. With these approaches, near duplicate social items can be repeatedly recommended to a certain user. Diversity-based approaches aim at mining a broad range of items that belong to different categories as diverse as possible and meanwhile, they are interesting to the target user. However, existing diversity-based approaches handle the user preferences as static, which ignores the temporal evolution of social users' preference. Recent recommendation approaches have been proposed to capture the user preferences over streams \cite{lommatzsch2015real,chandramouli2011streamrec,chang2017streaming,huang2016real}. They mainly focus on how to extend the traditional recommendation techniques such as matrix factorization \cite{he2016fast} to streaming environments by applying them to media data with the support of efficient stream processing. These approaches can efficiently conduct stream recommendation as they do not need to consider the whole user viewing history, which ignores the long-term interests of users and the requirements of broad item coverage to users. However, long-term interests reflect users' inherent characters and their stable preferences over life, which greatly affects users' behaviors in social activities. For example, John regularly enjoys movies online after work. Recently, affected by the war in Syria, John has browsed some videos related to this war. However, when the war is ended, John would get back to his regular activity of watching movies in spare time, and still hope to receive recommendation on movies.
Meanwhile, redundant items are added to user profiles, which is a barrier to the representation ability and visibility of their preferences.

In this paper, we propose a graphical model-based framework for effective and efficient social item recommendation over streams. Specifically, we first propose a novel Bi-Layer Hidden Markov Model (BiHMM) to capture each user' media browsing history and his interest patterns over a set of media producers for predicting his next interested media category. To measure the relevance between a user and an item, we design an entity-based item-user ranking function, which considers the short-term and long-term interests, and the diversity of the recommended items. Finally, we generate recommendation over streams based on the relevance between an incoming item and each user, and accelerate this process by using a novel signature-tree-based index scheme called CPPse-index. The main contributions of this work are summarized as follows:

\begin{itemize}
\item We propose a novel graphical model called Bi-Layer Hidden Markov Model (BiHMM) to predict users' long-term and short-term interests. BiHMM well captures users' interest dependency over various media producers.
\item We propose a novel item-user matching scheme that embeds the users' long-term and short-term interests, and the item descriptions over their expanded entities. The new matching scheme takes into account the diversity issue of recommendation.
\item We design a new CPPse-index scheme to improve the recommendation efficiency, which is guaranteed by a novel upper-bound-based candidate pruning.
The test results prove the effectiveness and efficiency of our approach.
\end{itemize}

The remainder of this paper is organised as follow. Section \ref{sec:related-work} reviews the related work on streaming recommendation. Section \ref{sec:pd} formulates our social media recommendation over streams. Section \ref{sec:hmm} presents our BiHMM model for user interest prediction, and our proposed matching scheme between items in media stream and social users, followed by our index scheme in Section \ref{sec:index}. We report the experimental evaluation results in Section \ref{sec:experiment}, and conclude the whole work in \ref{sec:conclusion}.


\section{Relate Work}
We review existing literature on two topics closely related to our work, including the recommendation over streams and the diversity-based recommendation.
\label{sec:related-work}
\subsection{Recommendation over streams}
Approaches have been proposed for recommendation over social streams \cite{chandramouli2011streamrec,diaz2012real,zhuang2013fast, chen2013terec,lommatzsch2015real,huang2015tencentrec,subbian2016recommendations,chang2017streaming}.
Most of stream recommender systems focus on adapting the traditional approaches to stream settings.
\citet{chandramouli2011streamrec} designed the StreamRec system, where the user-item interaction matrix for collaborative filtering is only updated when a subscription list is changed, which reduces the time cost greatly. The matrix factorization (MF) is the most popular technique in CF-based recommendation. However, it cannot be directly applied to stream-based recommendation due to the high cost of computing the stochastic gradient descent (SGD). To solve the problem, {\citet{zhuang2013fast}} proposed a parallel SGD which greatly speeds the SGD calculation.
\citet{diaz2012real} consider collaborative filtering as
an online ranking problem and present Stream Ranking Matrix
Factorization (RMFX) for optimizing the personalized
ranking of topics.
\citet{chen2013terec} models users and items using competitive matrix factorization for temporal stream recommendation.
\citet{lommatzsch2015real} apply the traditional collaborative filtering to the user interaction patterns within the recent time window. However, this technique is only applicable for items with strong temporal patterns, such as news articles.
{\citet{huang2015tencentrec}} conducts collaborative filtering over Apache Storm, which achieves high efficiency in stream recommendation. \citet{subbian2016recommendations} proposed a probabilistic neighbourhood-based
algorithm for performing recommendations in real-time. The similarity between a given item and each of all other
items is computed. The rating of a user to a particular item is predicted by calculating the weighted average of the ratings of its most similar items in this user's profile.
\citet{chang2017streaming} model user-item relationship with the temporal dynamics incorporating both hidden topic evolution and new user/item introduction. These
collaborative filtering-based approaches highly rely on the user ratings, which is not reliable over streams, thus the effectiveness of recommendation can not be guaranteed. In this work, we aim to solve the stream recommendation problem by predicting user long-term and short-term interests, constructing robust user models over them, and generating the recommendation results by matching user profiles and each incoming item.

\subsection{Diversity-based recommendation}
Traditional diversity-based recommender systems exploit the item-item relationship for achieving as diverse results as possible. Typical diversity-based recommendation
can be classified into two categories: (1) recommendation candidate re-ranking-based \cite{zhang2005improving,tong2011diversified,he2012gender,hurley2013personalised}; and (2) candidate filtering-based. Recommendation candidate re-ranking-based methods generate a list of recommendation candidates re-ranked based on the similarity between each other, such that the diverse results appear at the top ranked positions. \citet{zhang2005improving} maintain a list of recommendation candidates that are updated iteratively based on the PageRank scores of the candidates and the new items in data collection. \citet{tong2011diversified} and \citet{he2012gender} use greedy algorithms to select the diverse items such that the distance between the current selected item and its previous one is maximized. \citet{hurley2013personalised} ranks the items based on each of their attributes, and the overall ranks of these items are obtained by integrating the weighted pairwise rank difference. The core of re-ranking-based methods is to adjust the order of the resulting list. Thus the diversity of recommendation is limited to a small scope.

Candidate filtering-based approaches  directly exclude the items in data collection close to those in the resulting list in the recommendation generation to achieve the high diversity of the recommendation.
In \cite{yu2017accuracy}, the diversity is introduced by measuring the dissimilarity between items and the preference of the target user with respect to the item to select the items that are far from each other but well match users' preference.
In \cite{hurley2011novelty}, the trade-off between diversity and matching quality is formulated as a binary optimization problem, and the diversity level can be explicitly tuned. In \cite{ribeiro2015multiobjective}, the recommendation is treated as a multi-objective problem that combines several recommendation methods in a way of maximizing the diversity.
\citet{puthiya2016coverage} represent the items as a similarity graph, and conduct recommendation by finding a small set of unrated items that best covers a subset of items positively rated by the user.
These approaches do not consider the diversity in items themselves, which can provide more candidates in recommendation generation. The notations used in this paper are listed in Table \ref{fbl-notations}.

\begin{table}
	\centering
	\caption{Notation Table. }\label{fbl-notations}
\setlength{\tabcolsep}{0.5ex}
\begin{tabular}{ll}
\toprule
Notation &
	Definition \\
	\midrule
$C = \{c_1, c_2,\dots, c_N\}$
	& Pre-defined categories of social items.\\
$p(u)$ & The activity pattern of a social user    \\
$\up$
	&  Producer, the user that creates the item.  \\
$\uc$
	&  Consumer, the user who browsed the item.  \\
$E$
	& A set of extracted entities. \\
$v = \langle{c,\up, E}\rangle$
	& A social item. \\
$\mathcal{L}_i = \langle{v_0,v_1,\dots, v_n}\rangle$
	& The long-term interest list of  $\uci{i}$.\\
$\mathcal{W}_i = \langle{v_0,v_1,\dots, v_w}\rangle$
& The short-term interest window of  $\uci{i}$.\\
\bottomrule
\end{tabular}
\vspace{-2ex}
\end{table}


\section{FRAMEWORK OF OUR SOLUTION}\label{sec:pd}

\begin{figure}[t]
	\centering
	\includegraphics[scale=0.22]{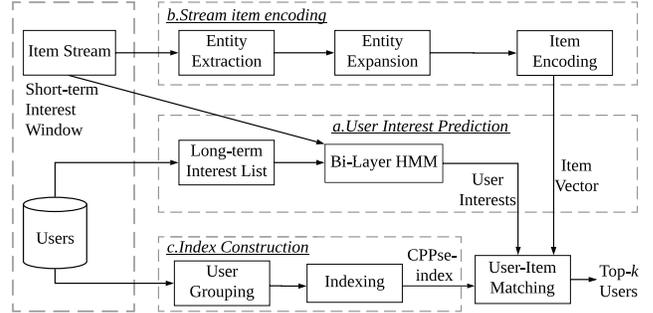}\vspace{-2ex}
	\caption{\small Framework of the stream recommendation}
	\label{fig:framework}\vspace{-3ex}
 \end{figure}

In this work, we propose a social stream and item stream
Recommendation framework (ssRec), as shown in Fig.~{\ref{fig:framework}}.
Our framework includes two major components, the user interest prediction and the user-item
matching. Besides, we design the {\Tree} to optimize the efficiency.
The user interest prediction predicts users' interests based on Bi-layer HMM (BiHMM)
model, as shown in Fig.~{\ref{fig:framework}}(a).
The user-item matching provides a ranking function between a stream item and a social user
based on the predicted interests.  Given a stream item, we encode it as an item vector as shown in
Fig.~{\ref{fig:framework}}(b), which is further used for the matching between the item and user profiles.
We propose a novel index structure, {\Tree}, to facilitate the recommendation process as shown in
Fig.~{\ref{fig:framework}}(c) . Users with the same interest are grouped together. We output top ranked
users by searching the {\Tree}. We will detail these modules in Sections~\ref{sec:hmm}-\ref{sec:index}.

\section{Bi-Layer HMM-Based Recommendation Model}
\label{sec:hmm}
We will present our Bi-Layer HMM (BiHMM) that predicts the category which a user may
browse immediately after the current time, and a probability-based item-user ranking.

\begin{figure}[t] 
	\centering
	\includegraphics[scale=0.7]
    {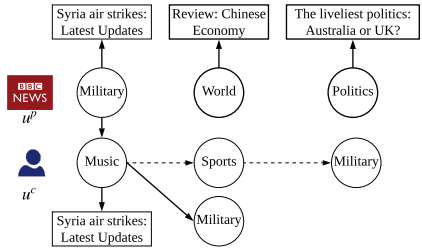}
	\vspace{-2ex} \caption{The application scenario, where the $u^c$ is a user browsing
		content created from the BBC news, which is a producer $\up$.}
	\label{fig:hmm-example}\vspace{-1.5ex}
\end{figure}

\subsection{The Bi-Layer HMM Model}
An important feature of social platforms is the user engagement. Users can create new social items instead of  consuming media only.
Thus, in this work, we consider a user in two modes: (i)  the {\Oc} and (ii) the {\User}:
\begin{definition}\label{def-user-modes}
	A user creating social items is a  \emph{producer} ($\up$), and
	a user browsing social items is a \emph{consumer} ($\uc$).
\end{definition}
\noindent Note that a user can be either a {\Oc} or a {\User}. Users who are only in {\Oc} mode like BBC News are
regarded as data sources, and do not receive any recommendations.

\begin{figure}[t]
	\centering
	\includegraphics[scale=0.6]{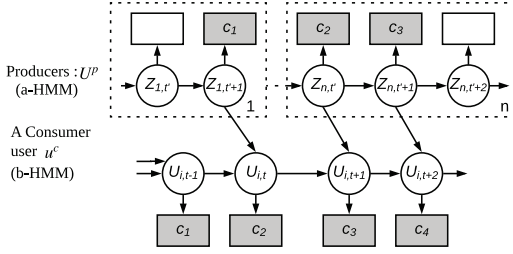}\vspace{-1ex}
	\caption{The BiHMM model. $Z_{i,t'}$ is the
		$t'$-th Hidden state of the influential user $i$, $U_{i,t}$ is $t$-th
		Hidden state of the producer $u^p_i$, and $c_i$ is the  item category.}
	\label{fig:hmm}\vspace{-4ex}
\end{figure}

Consider a real scenario shown in Fig.~{\ref{fig:hmm-example}}. A user behavioral trajectory may follow the categorical pattern:
``music, sports and military''.
Such pattern may also exist in the social item creating process. For example, BBC news may create social items following the
categorical pattern ``military, world and politics''.
Assuming a consumer's behavior is independent of the producer may be too strong to
be  applied in  real production systems.
As shown in the example, when a bursting event
happens and is captured by a $\up$ that a user is following, the regular
behavioral trajectory of the user is highly likely to be interrupted.
To capture this dependencies, we propose a Bi-Layer HMM as shown in Fig.~{\ref{fig:hmm}}.

Unlike the single-layer HMM that considers consumers' behavior only,
there are two layers in our model: a-HMM and b-HMM. The a-HMM layer captures the patterns on a set of producers that a user consumer $u^c$ is interested in, while the  b-HMM layer models his browsing trajectories.
Each dashed box represents one producer $\up$, circles are the current hidden states, and the gray rectangles represent observed behaviors. We use arrows to show the relation between two states. For example, if there is an arrow from $Z_{1,t'+1}$ to $U_{i,t}$, it means $U_{i,t}$ is decided by $Z_{1,t'+1}$.
Let $Z_{i,t}$ be the hidden state of a $u^{p}_i$ at time $t$ and $U_{i,t}$ be the hidden state
of a consumer user $\uci{i}$ at time $t$.
As a user's next state may be correlated with any of the $\up \in U^{p}$, the hidden states in both layers are dependent.
For example, if {\uci{i}} browsed an item in category $c_1$ under state $U_{i,t-1}$. and $c_1$ is
created by a producer $u^p_1$ under the hidden state  $Z_{1, t'+1}$, then under our model
setting, the next state $U_{i,t}$ of $\uci{i}$ is both  decided by $Z_{1,t'+1}$ and the current state $U_{i,t-1}$. We will discuss elements in the BiHMM model as follows.

\myparagraph{The a-HMM Layer for Modelling Producers}
We first build the a-HMM layer to model users that create social items. Assume that the activity of a user creating a social item is independent of other users. Then,
we can apply classic HMM technique to model the social item creation process for all producers. In the modelling process, three components need to be estimated:
(i) the hidden states $Z_{i}, (1\leq i \leq N^{(a)})$, where $N^{(a)}$ is the number of hidden states; (ii) the ${N^{(a)}\times N^{(a)}}$ state transition probability matrix $A^{(a)}$;
and (iii) the observation matrix $B^{(a)}$. Each element in  $A^{(a)}$ is computed using $a_{ij} = p(Z_j | Z_i)$, and each
element in $B^{(a)}$ is computed using $b_{jm} =p(c_m|Z_j)$.
Note that $\sum_{m=1}^{M}b_{jm}=1$, where $M$ is the number of observations.
Suppose that the initial state probability distribution is $\pi^{(a)}=\{\pi^{(a)}_i\}$,
where $1\leq i\leq N^{(a)}$, $\pi^{(a)}_i=p(Z_i),$ and 	$\sum_{i=1}^{N^{(a)}}\pi^{(a)}_i=1$. Based on our previous analysis, the parametrization of a-HMM is  $\lambda^{(a)}=\langle\pi^{(a)}, A^{(a)}, B^{(a)}\rangle$.
We use Baum-Welch algorithm \cite{welch2003hidden} to learn all three parameters.
In the prediction, given an observed category $c$, its associated hidden state is
obtained using Viterbi Algorithm~\cite{forney2005viterbi}.

\myparagraph{The b-HMM Layer for Modelling Consumers}
As emphasized, the interaction of a consumer to a social item depends on both the historical
trajectory of the user and that of the producers interest this consumer. Thus, we build the b-HMM by considering both the trajectory of a user's historical activities and that of its interactions with different producers.

Just as a-HMM, b-HMM has three major components: hidden states, state transition probability matrix and the observation matrix.
Let $N^{(b)}$ be the number of hidden states in b-HMM, $U_{i}$ the $i$-th hidden states.
Each entry $a_{ikj}^{(b)}$ in the  state transition matrix $A^{(b)}$ is then computed as
 $a_{ikj}=p(U_j|(U_i,Z_k))$, where $Z_k$ is the hidden state from producers.
Similarly, each entry $b_{jkm}$  in the observation probability matrix $B^{(b)}$ can be
obtained via $b_{jkm}=p(c_m|(U_j,Z_k))$.
Note that $c_m$ is the observed social item category,
Like a-HMM, the parametrized representation of b-HMM is  $\lambda^{(b)}=\langle\pi^{(b)}, A^{(b)},
B^{(b)}\rangle$, where $\pi^{(b)}$ is the initial state probability distribution and $\pi^{(b)}=\{\pi^{(b)}_i\} = \{p(U_i)\}$.

The classic parameter estimation approach for HMM cannot be directly applied to the estimation in b-HMM due to the dependency of its states to the a-HMM. Thus we reformulate the representation of b-HMM by integrating the states of two layers in the BiHMM.
Consider the next hidden state in b-HMM determined by both $U_i$ and $Z_k$ from
b-HMM and a-HMM respectively. We can denote the new state of b-HMM as $U' =\{U_i,Z_k\} $.
Accordingly, the state transition probability matrix $A^{(b)}$ can be converted into $A'^{(b)}=\{a_{(i\times k)j}\}=\{p((U_i,Z_k)|U_j)\}$. The observation probability matrix becomes $B'^{(b)}=\{b_{(i\times k)m}^{(b)}\}=p(c_m|(U_i,Z_k))$ after transformation.
The b-HMM is represented as $\lambda^{(b)'}=\langle\pi^{(b)}, A^{(b)'}, B^{(b)'}\rangle$.
Based on our new representation, we can train the b-HMM by the same way used in the a-HMM.

With the learned b-HMM, the next observation is predicted as follows. Given a series of observations $o=\{c_1,...,c_n\}$, we first predict the series of hidden states, $U'=\{U'_1,...,U'_n\}$, which have the highest probabilities of generating these observations. Then, we
exploit the Viterbi Algorithm \cite{forney2005viterbi} to predicate the top-$k$ categories interesting a user.

\subsection{Modelling User Profiles and Stream Data}
\myparagraph{Stream Data Models}
In our recommendation scenario, two types of data streams should be considered: the social item
data stream and the user-item interaction data stream.
The social item stream is generated as a data stream by the high-velocity media data uploading.
Let $V=(v_{0}, v_{1}, \dots, v_{t})$ be the social item stream over a time period $t$. We need to construct a model that well captures the items' content and contexts within the time window. Meanwhile, as the item uploading and users' interactions with items, the temporally frequent user-item interactions form a user-item interaction stream. A good data model should capture the user-item interactions in a streaming mode, rather than the static ones appearing in traditional recommender systems \cite{he2016fast,yao2015unified}.
In addition to the social property (the producer of the item), we also consider
the item itself, and specifically, entities in the item description. Given an item
$v$, we describe it as a triplet $\{c, \up,$ $ E \}$, where c is its category, $\up$ its producer, $E$ the set of extracted
entities from $v$.
Consider a video description as below:
\begin{quote}
 \textit{\underline{Australian Open} 2017 Men's Final
\underline{Roger Federer} vs \underline{Rafael Nadal} Full
\underline{Match}}.
\end{quote}
We can represent it as a set of its entities $E=$\{``Australian Open'', ``Roger Federer'', ``Rafael Nadal'', ``Match''\}.
Clearly, if a user's long-term interest list contains one or more of these entities multiple times in the category $c$, then it is highly likely
that the user would like the current social item $v$. However, considering the entities in an item only may generate the less diversified
recommendation results.
To solve this problem, we apply the expansion techniques on each entity.
Expansion entity sets are extracted based on the proximity heuristics \cite{tao2007exploration}, from item descriptions.
If two entities often co-occurred closely in the same category, we believe they are
strongly related.
Given two entities, the expansion weight between them is calculated by their proximity. Since we
consider all the entities in media descriptions, the location information on videos appearing as
entities is embedded in the model.

\myparagraph{User Models}
Similar to all classic recommender systems, we consider users' {\em long-term} interests,
which can be inferred from users' historical interaction logs. On the other hand, due to the
effect of some external events, users' interests may be changed in a short-term time period.
For example, users who usually watch sport news only may start following some political news as the poll commences. Thus, we consider the {\em short-term} interests of users as well. Both long-term and short-term interests are important. If we only consider the long-term interests, the recommendation results lose the
{\em recency}. Reversely, considering the users recent activities only will lead to the users {\em interest
drift}.

For each user, we construct a user profile based on the long-term interest list  and short-term interest window. Instead of tracking fine-graind social items, we consider their categories only, as it is enough for us to infer users' interest patterns through item categories. The short-term interest window of a user has a fixed-size, and keeps his latest interaction records, while his long-term interest list includes all the rest of records in his whole browsing history.
Let $\mathcal{L}_i$ be the long-term interest list of a user $u_i$, which is a social item sequence in temporal order.
If we consider each item $v$ using a pair $\langle$category-\Oc$\rangle$, then $\mathcal{L}_i=\left(\langle{c_0, u_1}\rangle, \langle{c_0, u_2}\rangle, \langle{c_2,u_1}\rangle \dots \langle{c_i, u_n}\rangle \right)$.
We maintain users' short-term interests $\mathcal{W}_i$ within a fixed-size recent time window in the way similar to  $\mathcal{L}_i$. When the short-term interest window is full,  $\mathcal{W}_i$ will be flushed to $\mathcal{L}_i$.
As such, each user profile is modelled as a pair of category-producer sequences (CPPse), which describes the long-term and short-term user interaction patterns.

\subsection{Entity-Based Item-User Matching}
\label{sec:streamuser-match}

Using our BiHMM model, we can compute the probability
of a media consumer $u^{c}$ browsing a specific category $c$. However, our ultimate goal
is to identify the relevance score between a user and an item. Thus, we need to design a user-item matching based on the output of BiHMM.

We calculate the relevance score of an item $v$ to a user over his long-term interests $u^{c}$ by
estimating the probability of $v$ matching $u^{c}$, denoted as $p (u^{c} | v )$, as below:
\begin{equation}\label{eqn-long-term}
\begin{aligned}
R_{\ell}(v, u^{c}) {} \propto~& p( u^{c} |v ) {p} ( u^{c} |\langle{c, \up, E}\rangle)  \\
\propto~&  {p}(c | u^c)\cdot  \hat{p}(\up| u^c) \cdot\prod_{e\in E\cup E'}w_e\cdot \hat{p}(e| u^c)
\end{aligned}
\end{equation}
in which $p(c|u^c)$ is the probability output by the BiHMM that describes the likelihood
of a user browsing an item in category $c$;
 $w_e$ is the  expansion weight if the entity $e$ is from the expansion set $E'$, otherwise $w_e=1$.
In our solution, we apply proximity heuristics to compute the expansion weights, considering the
co-occurrences of entity pairs.
Both $\hat{p}(\up|u^c)$ and $\hat{p}(E|u^c)$ are
estimated using  Maximum  Likelihood Estimation (MLE), indicating the probability of a user being interested in the item given the producer and the
description respectively.
For computational convenience, we consider the log-likelihood score and reformulate the long-term-based recommendation score computation as:
\begin{equation}
\label{eq-ranking-model}
\begin{aligned}
R_{\ell}(v, u^{c}) & =  \log {p}(c|u^c) + \log \hat{p}(\up|u^c) \\									&+ \log\sum_{e\in E \cup E'} w_e\cdot  \hat{p} (e|u^c)
\end{aligned}
\end{equation}

It is known that the entities and producer of the current item may have never
appeared in the user's long-term history. Under this situation, a zero probability will be
given in the MLE estimation.
This may hamper the effectiveness on the diversification and serendipity.
To prevent the zero probability, we apply the Dirichlet smoothing technique to both producer and
entities.
The final recommendation score is parametrized by $\lambda_s \in (0,1)$, integrating
long-term and short-term scores:
\begin{equation}\label{eqn-rec-score-combined}
R(v, u^{c}) = (1-\lambda_{s}) \cdot  R_{\ell}(v, u^{c})+\lambda_{s} \cdot R_{s}(v, u^{c}),
\end{equation}
in which $R_{\ell}(v, u^{c})$ is computed using Equation~{\ref{eqn-long-term}}, and
$R_s$ is the score computed using the same function but based on users' short-term interests:
\begin{equation}\label{eqn-score-short}
 R_s(v,\uc)=\log {p}(c |u^c).
\end{equation}
Note that for the short-term interest, we only consider the prediction probability output from the BiHMM model.
This is because we only maintain a window of recent items and the MLE estimation over a few social items leads to the imprecise estimation results on the user interests.

\section{Recommendation Generation Optimization}\label{sec:index}
We present our recommendation generation in details. Given an incoming social item $v$, a collection of social users $\{u^c\}$, and a relevance function $R$, our recommendation finds a list of social users $S$ with the best relevance to $v$, i.e., for any $u_1^c \in S$ and $u_2^c \notin
S$, $R(v, u_1^c) \geq R(v, u_2^c)$ holds. To perform the recommendation, a naive method is to
compute the similarity between $v$ and each of social users. Given a set of $n$ users,
this naive method requires $n$ relevance calculations, which is inappropriate to high speed
streams. High-dimensional indexes, like R-tree variants \cite{Beckmann:1990:RER:93597.98741}, and B$^+$-tree based indexes \cite{DBLP:journals/tkde/ZhouZCSBT10, DBLP:journals/vldb/ZhouZCB12}, are
not designed for online processing, thus inapplicable to our problem either.
An efficient index scheme is demanded for the online environment.

\subsection{The CPPse-index structure}\label{sec:indexstructure}
\begin{figure}[t]
	\centering
	\includegraphics[scale=0.3]{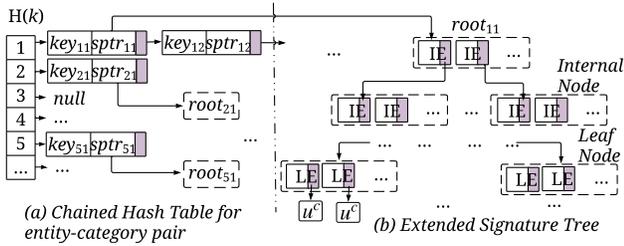}
	\caption{\small CPPse-index structure. Shaded rectangles represent
		pointers, and $\uc$ at the bottom represents user profile records. Note that both
		IEntry (IE) and LEntry (LE) contain pointers.}\vspace{-3ex}
	\label{fig:tree}
\end{figure}
We propose the {\Tree} to improve  recommendation efficiency.
Our index includes two core parts: (1) a chained hash table that maps each online item to its
extended signature-trees; (2) a number of extended signature-trees, each of which stores  user
profiles over a particular category in a user block. User blocks are generated by one pass clustering \cite{Schweikardt2009}
based on each user's long-term categorical interests and cosine similarity.
We construct an extended signature tree for each category of a user block.
As such, the number of  entities covered by a signature is greatly reduced, and the signature
representation is highly  compact, leading to a compact signature tree. Fig.~\ref{fig:tree} shows
our {\Tree} structure.

We use chained hash tables to
organize the category-entity pairs due to its simplicity. Like \citet{zhou2015online}, we select
the class of \textit{shift-add-xor} string hashing functions for mapping category-entity pairs to hashcodes,
considering their important properties such as uniformity, universality, applicability and
efficiency \cite{DBLP:conf/dasfaa/RamakrishnaZ97}. Let $s=c_1,...c_m$ be a string of $m$
characters, $s$ a seed and $h_i$ an intermediate hash value after examination of $i$ characters.
The components in the class of \textit{shift-add-xor} are defined as:
\begin{equation}
\begin{array}{r}
 \var{init}(s)=s ~~~~~~~~~~~~~~~~~~~~~~~~~~~~~~~~~~~~~~~~(a) \\
 \var{step}(i,h,c)  =h\bigoplus(\mathbf{L}_L(h)+\mathbf{R}_R(h)+c)~~~~~~~~~(b) \\
 \var{final}(h,s)   =h||T~~~~~~~~~~~~~~~~~~~~~~~~~~~~~~~~~~~~(c)
 \end{array}
\label{shiftAddXor}
\end{equation}

\noindent Here, $\mathbf{L}_L(h)$ denotes the left-shift of value $h$ by $L$ bits,
$\mathbf{R}_R(h)$ is the right-shift of value $h$ by $L$ bits. Given a pair of item category and
entity that forms a phrase, we first generate an initial hash code using the equation
\ref{shiftAddXor} (a), then recursively compute the intermediate hash code over its first $i$
characters using the equation \ref{shiftAddXor} (b), and finally obtain the modulo value of the
hash code over its $m$ characters. Given a set of  category-entity names, we organize it as a
chained hash table with a number of hash buckets. Each element of the hash table is a triad
denoted as  $\langle \var{key}, \var{sptr}, \var{nextptr}\rangle$, where \textit{key} is the
hash value,  \textit{sptr} the set of pointers to the extended signature-trees under
this category, and  \textit{nextptr} pointer to the next category-entity pair with the same
hash code.
Given a category-entity pair, its hash bucket is located based on its hash code, and its triad is
inserted into this bucket. A chained hash table is constructed by inserting the triads of all the
category-entity pairs in the media set. Each category-entity pair can be at most covered by $|B|$
user blocks, so at most $|B|$ \textit{sptr} are needed, where $|B|$ is the total amount of user
blocks. If an extended signature-tree does not contain  this category-entity pair, the
corresponding pointer will point to null.

Signature-tree is a high-dimensional index structure derived from  R-tree family, with more
efficient querying and updating. Such improvement is obtained by generating bitmap encoding
and then perform boolean conjunction queries. Apparently, it is not applicable in our setting as
we need to consider the quantification of social consumer activities while encoding its signatures.
To handle this problem, we extend the  signature-trees by designing a new encoding scheme: an
impact encoding for maintaining user profiles and a frequency-based encoding for queries.
We construct two types of entries in the extended signature-tree: an internal entry
(IEntry) that summarizes statistics of its children and a leaf entry (LEntry) that
represents a user's profile. Given a user $\uc$ in  block $B$ under a category $c_i$, the leaf
entry contains the user's long-term interest list and short-term interest window. Its long-term
interest list is described as a tuple  $\mathcal{L}_{u | c_i} = \langle{p_{\ell}(c_i), P_{U^{p}|
c_i},|U^p|, P_{E| c_i}, |E| }\rangle$, where $p(c_i)$ is the probability of user browsing an item
in $c_i$, $|U^p|$ and $|E|$ are total numbers of producers and entities in the users' history,
respectively. The $P_{U^{p}|c_i}$ and  $P_{E| c_i}$ are two impact lists, storing lists of
$\hat{p}(\up | u)$ and $\hat{p}(e | u)$, respectively.
Users' short-time interest representation is constructed from the most recent item sequence, stored in a fixed-length window.
In addition to the user profile statistics,  each LEntry is also attached with a pointer to its user profile record.

IEntry is created during the construction of a tree. An IEntry is a virtual ``user'' whose interests cover all of its children. Like LEntry, data statistics on the long-term and short-term interests of the virtual ``user'' are extracted as the signature of this IEntry, which are computed  by applying $\max()$ to all children over their corresponding signature components. Similarly, an IEntry is attached with a pointer to its child subtree.

\begin{table}[t]\small
	\centering\caption{The factors relevant to user profile signature size}\label{tab:usersigsize}\vspace{-1ex}
	\begin{tabular}{l c c c c c c c }
		\toprule
		User block num & 1    &10  & 20  & 30 & 40 & 50
		\\[0.2ex]
		Max entity num  & 4000 &475 & 257 & 155 & 123 & 101
		\\[0.2ex]
		Max producer num & 98  &53 & 40 & 39 & 32 & 25
		\\\bottomrule
	\end{tabular}\label{tab:statistics}\vspace{-4ex}
\end{table}

Table~\ref{tab:statistics} shows the statistics over our Youtube dataset. As we can see, applying user blocking reduces the entry size in a tree by large. Without blocking, 4{,}000 entities should be considered for each entry, which is infeasible for the memory-based index. The size of entity set over a user block is much smaller, which greatly saves the memory cost of the index.

\subsection{KNN query}\label{sec:query}
We perform the item-user matching by top-$k$ query over the {\Tree}. Before we proceed to the detailed KNN query algorithm, we need to decide how to generate a pseudo-query given an item and an extended signature-tree, and how to measure the relevance score of an item to an IEntry.

As introduced previously, each incoming item is described as a triplet $v=\langle{c, \up,
E}\rangle$, where $c$ is the category of the social item, $\up$ is the creator of the social item
and $E$ is the set of entities extracted from the social item. However, the triplet representation
cannot be directly used as a query. To conduct KNN query over {\Tree}, we need to generate a
signature for the item.
Given an item $v$, let $c$ be its category, its pseudo-query is generated by converting its $\up$ to one-hot encoding over $U_i^p$, and entity set $E_v$ into the frequency vector over the entity set $E_i$ for each block containing $c$ and any entities in $E_v$, where $i$ is user block $id$. We also keep a $|E_i|$-dimensional vector to record the weight of each entity. Note that, if the subtrees constructed from $n$ different user blocks are identified based on the category $c$ and the entity set $E_v$ of item $v$, then $n$ different encodings will be generated for $v$. The following example illustrates the query
generation process.
\\[1ex]
\noi\textbf{Example 1.}
Suppose we have an incoming item $v$ in category {\em sports} with a collection of entities $E$. Let $\{B_0, B_1,B_2\}$ be 3 user blocks containing category {\em sports} and some elements in $E$.
We have	$B_0=\langle{U_0^p,E_0}\rangle$, where
	$U_0^p$=\{weSpeakFootball,
	{Wrzzer}, {SirMan},
	{bundesteam}\}, $E_0=$ \{{Beckham}, {football}, {worldcup}, {FIFA}, {Brazil}, {Messi}\}.
	Suppose $v=\langle{c, \up, E}\rangle$,
	$\up$ is Wrzzer, and {$E=\langle$Beckham}, {worldcup}, {worldcup$\rangle$}. After expansion, the entity set of $v$ becomes $E'$={$\langle$Beckham, Messi, worldcup, FIFA, worldcup, FIFA$\rangle$}, and the entity weight vector of $E'$ is $\langle 1,0.7,1,0.9,1,0.9\rangle$. We encode all elements of $v$, generating the signature of $u^p$ over $U_0^p$ $\langle 0,1,0,0\rangle$, the signature of $E'$ over $E_0$ $\langle 1,0,2,2,0,1\rangle$ together with its weight
	vector $\langle 1,0,1,0.9,0,0.7\rangle$. The two signatures and the entity weight vector are connected to form a complete signature of item $v$. Thus the signature of $v$ over	$B_0$ is $q_{v,0} =\{0,\var{sports},\langle 0,1,0,0\rangle,\langle 1,0,2,2,0,1\rangle,\langle	1,0,1,0.9,0,0.7\rangle\}$. By the same way, we can generate the signature of $v$ over $B_1$ $q_{v,1} =\{1,sport,\langle
0,1\rangle,\langle 1,0,3,2\rangle,\langle 1,0,0.7,0.6\rangle\}$, and that over $B_2$ $q_{v,2}
=\{2,sport,\langle 1,0\rangle,\langle 1,3\rangle,\langle 1,0.8\rangle\}$.

Given an item $v$ and an IEntry, we define their relevance function, which is the {\em Recommendation Upper Bound} of the measure between $v$ and an LEntry below this IEntry.

\begin{definition}\label{def-rec-ub}
	Consider an internal entry $\mbox{IEntry}=\{p_{\ell}(c), P_{U^p},P_E, p_{s}(c)\}$ and an item
	$v=\langle{c,\up, E}\rangle$.
	After encoding the item $v$ into $q_v=\langle{c,F^v_{U^p}, F^v_E}\rangle$,
	the relevance between them can be computed by plugging statistics into
	Equation~{\ref{eqn-rec-score-combined}}, which then becomes:
	\begin{equation}\label{eq-ranking-model}
\begin{aligned}
	R(v,\mbox{IEntry})
	& =  (1-\lambda_{s}) \cdot ( \log {p}_{\ell}
	 +\log{F^v_{U^p}\cdot P_{U^p}}\\ &+ \log { F^v_E \cdot ( W_e \otimes P_E) })+ \lambda_{s} \cdot \log {p}_{s},
\end{aligned}
\end{equation}
where  $p_\ell$ and  $p_s$ are  maximal BiHMM probability of all I-Node's children, for the long-term and
short-term interests, respectively;
$E$ is a set of entities kept in the current tree,
$W_e$ is the expansion weight vector corresponding to the entity. According to Definition~{\ref{def-rec-ub}}, we have the following lemma:
\end{definition}

\begin{lemma}\label{lemma-ub1}
Given an internal entry IEntry and an item $v$, for any internal entry IEntry' in the subtree of IEntry, the following inequality holds:
	$$R(\mbox{IEntry}, v) \geq R(\mbox{IEntry}', v) \,$$
\end{lemma}
\noi\textit{Proof.}
Suppose $\mbox{IEntry}=\{p_{\ell}(c), P_{U^p},P_E, p_{s}(c)\}$, $\mbox{IEntry}'=\{p'_{\ell}(c), P'_{U^p},P'_E, p'_{s}(c)\}$. From the construction of CPPse-index, we know $p_{\ell}(c) \geq p'_{\ell}(c)$,
$\forall i~P_{U^p}(i) \geq P'_{U^p}(i)$, $\forall j~P_E(j) \geq P'_E(j)$ and $p_s(c) \geq p'_s(c)$.
Suppose the item $v$ is encoded as $q_v=\langle{c,F^v_{U^p}, F^v_E}\rangle$. Given $F^v_{U^p}$, as $P_{U^p}(i) \geq P'_{U^p}(i)$, we have:
\[
\sum_i F^v_{U^p}(i)\cdot P_{U^p}(i)\geq \sum_i F^v_{U^p}(i)\cdot P'_{U^p}(i).
\]
Thus, we have $F^v_{U^p}\cdot P_{U^p}\geq F^v_{U^p}\cdot P'_{U^p}$. As logarithmic is strictly
monotone increasing, $\log F^v_{U^p}\cdot P_{U^p}\geq \log F^v_{U^p}\cdot P'_{U^p}$. By the same
way, we have $\log{P^v_E \cdot ( W_e \otimes P_E)}\geq \log{P^v_E \cdot ( W_e \otimes P'_E)}$.
Thus, $(1-\lambda_{s}) \cdot ( \log {p}_{\ell} + \log{F^v_{U^p}\cdot P_{U^p}} + \log { F^v_E \cdot ( W_e \otimes P_E) })+ \lambda_{s} \cdot \log {p}_{s} \geq (1-\lambda_{s}) \cdot ( \log {p'}_{\ell} + \log{F^v_{U'^p}\cdot P_{U'^p}} + \log { F^v_E \cdot ( W_e \otimes P'_E) })+ \lambda_{s} \cdot \log {p'}_{s}$, i.e.,
\[
R(\mbox{IEntry}, v) \geq R(\mbox{IEntry}', v). \qquad \qed
\]

\begin{lemma}\label{lemma-ub}
Given an internal entry IEntry and an item $v$, for any user $\uc$ covered by I-node,	the following inequality holds: $$R(\mbox{IEntry}, v) \geq R(\uc, v)\,$$
\end{lemma}
\noi\textit{Proof.}
Suppose $\mbox{IEntry}=\{p_{\ell}(c), P_{U^p},P_E, p_{s}(c)\}$ and $u^c=\{p''_{\ell}(c), P''_{U^p},P''_E, p''_{s}(c)\}$. $v$ can be encoded as $q_v=\langle{c,F^v_{U^p}, F^v_E}\rangle$.
We have $p_{\ell}(c) \geq p''_{\ell}(c)$,
$\forall i~P_{U^p}(i) \geq P''_{U^p}(i)$, $\forall j~P_E(i) \geq P''_E(i)$ and $p_s(c) \geq p''_s(c)$.
We consider two cases.
\begin{itemize}
\item If IEntry is the parent of item $v$, we can directly have
$R(\mbox{IEntry}, v) \geq R(u^c, v)$.

\item If IEntry is not the parent of $v$, we can find a branch in our CPPse-index that from IEntry
to $v$, say IEntry, $N_0$,...,$N_n$, $v$. By Lemma \ref{lemma-ub1}, we have
$R(Ientry,v)>R(N_0)>...R(N_n,v)$. By i), we have $R(N_n,v)>R(u^c,v)$. Thus, $R(\mbox{IEntry}, v)
\geq R(u^c, v)$. \qed
\end{itemize}

\begin{algorithm}[t]
	\caption{KNN Query Processing}
	\label{alg:topk}
\begin{algorithmic}[1]
	\State \textbf{Input}: CPPse-index and $v=\langle{c,\up, E}\rangle$ the social
	item
	\State \textbf{Output}: a ranked list of users, $U_{k}$
    \State $T \gets \{\},\,Q\gets \{\},\,U_k\gets \{\}$
	\Comment{$U_{k}$ is a size $k$ max-heap}
    \For{each $e$ in $E$}
    \label{stp-get-preprocess-start}
    \State  $\var{key}  \gets \func{calc\_hash}(e, c)$
    \label{stp-get-tree-start}
    \State  $\var{ptr} \gets \func{get\_tree}(\var{key})$
     \label{stp-get-tree-end}
    \Comment{{\em ptr} is a pointer to current tree}
    \State $q_v \gets \func{gen\_pusedo\_qry}(v, \var{ptr})$
    \label{stp-query-gen}
    \Comment{encoding $v$ w.r.t { tree}}
    \State $T \gets T \cup \var{tree},\,Q\gets Q\cup q_v$
    \EndFor
    \label{stp-get-preprocess-start}
	\State $\var{curr\_p} \gets \{\},\,\var{LB}\gets 0$	
    \Comment{{\em curr\_p} is a priority queue}
    \For{$i$ \textbf{from} 0 \textbf{to} $|T|-1$}
    \For{all {\em entry} in  node that $\var{ptr}$ points to}
    	\State \func{Enqueue}$(\var{curr\_p},\langle{R(Q_i, \var{entry}), \var{entry},Q_i}\rangle)$
     \EndFor
    \EndFor
    \While{\var{curr\_p} is non-empty}
    \label{stp-ranking-start}
    \State $\langle{\var{score} , \var{entry}, q_v} \rangle \gets \func{Dequeue}(\var{curr\_p})$
    \If{$score > \var{LB}$}
    \label{stp-score-lb-cmp}
    \If{\var{entry} is LEntry}
    \State $\var{LB} \gets \func{Insert}(U_k)$
    \label{stp-LEntry-start}
    \Comment{Update heap}
    \Else
    \For{all {\em c\_entry} in node that $\var{entry}$ points to}
    \label{stp-IEntry-start}
    \State $\var{score} \gets R(q_v, \var{c\_entry})$
    \If{$\var{score} > \var{LB}$}
    \State  \func{Enqueue}$(\var{curr\_p},\langle{\var{score} ), \var{c\_entry}, q_v}\rangle)$
    \EndIf
    \label{stp-IEntry-end}
    \EndFor
    \EndIf
    \EndIf
    \EndWhile
    \label{stp-ranking-end}
    \State {\bf return} $U_k$ \label{al:topkz}
\end{algorithmic}
\end{algorithm}
\setlength{\textfloatsep}{0.8pt}

Lemmas \ref{lemma-ub1}-\ref{lemma-ub} guarantee no false item pruning can happen in the query processing. Algorithm \ref{alg:topk} illustrates the general framework for the KNN query over CPPse-index. Given an incoming social item $v$, our algorithm performs KNN query by
three important steps:
(1) compute the hash values based on the entity--category pairs contained in $v$, by which a set of extended signature trees are located (Lines~\ref{stp-get-tree-start}-\ref{stp-get-tree-end});
(2) generate pseudo-query based on the item and each located extended signature tree  (Line~\ref{stp-query-gen});
(3) select and rank the top-$k$ relevant users (Lines~\ref{stp-ranking-start}-\ref{stp-ranking-end}).
We maintain a max-heap $U_k$ with size $k$ as our output ranked list.
In the ranking process, a priority queue $\var{curr\_p}$ is maintained, including the recommendation score to the entry, current entry and the generated query. The recommendation score is used as the comparison key to decide if the priority queue should be updated. The queue will be updated if the current recommendation score is bigger than the lowest score kept in resulting heap $U_k$.
In the entire process, we only consider the entries that have scores larger than $\var{LB}$.
When the current entry is an IEntry, its children will be put into the priority queue (Lines~\ref{stp-IEntry-start}--\ref{stp-IEntry-end}) if their score is larger than the $\var{LB}$ in $U_k$; otherwise we update the heap $U_k$ (Line~\ref{stp-LEntry-start}).

\subsection{Dynamic Maintenance}
\label{sec:update}

This section discusses the dynamic maintenance of our CPPse-index. In social communities, the user information is highly dynamic due to the frequent user activities. For one thing, when users browse media data, their user interest patterns change. Users may browse the media containing existing entities, which changes the entity frequencies in their user profile signatures. Users may browse the media covering new incoming entities as well, which expands their signatures and adds new category-entity pairs to be kept in the hash table. For another, new users may join social community, which adds new profiles to be maintained. We need to maintain the short-term interest window, update the user profile representations and all their ancestor internal entries in CPPse-index to reflect the recent updates in social community.
\begin{algorithm}[t!]
\caption{User Profile Update Maintenance}
\label{alg:update}
\begin{algorithmic}[1]
	\State\textbf{Input}:
	{\Tree} and $\{u^c\}$, user profiles to  update
	\State \textbf{return} updated {\Tree}
	\For{each $\uc$ in \{\uc\}}
	\State $\func{UpdateUserProfileRepresentation}(\uc)$
	\label{stp-locate-tree-start}
	\For{all $e, c$ pairs in $\uc$ history}
	\State $\langle{\var{keys},\{\langle{c, \var{ne}}\rangle\}}
	\rangle \gets \func{calc\_hash}(e,c)$
	\State\Comment{$\{\langle{c, \var{ne}}\rangle\}$ is a set of new entity category pairs}
	\EndFor
	\State $\{\var{ptrs}\} \gets \func{get\_tree}(\var{keys})$
		\label{stp-locate-tree-end}
	\Comment{get all extended trees}
	\State Insert $\{\langle{c, \var{ne}}\rangle\}$ to hash table if it is non-empty.
	\label{stp-hash-update}
	\For{each $\var{ptr}$ in $\{\var{ptrs}\}$}
	\State $\var{LE} \gets \func{find\_leaf\_entry}(\var{ptr}, \uc)$
	\label{stp-find-le}
	\If{find {\em LE}} update {\em LE} and its ancestors
	\label{stp-update-prof1}
	\Else~$\func{insert\_to\_index}(\uc)$
		\label{stp-update-prof2}
	\EndIf
	\EndFor
	\EndFor
	\State \textbf{return} \Tree
\end{algorithmic}
\end{algorithm}

We maintain the CPPse-index periodically by checking the activities of social users.
Algorithm \ref{alg:update} shows the detailed process of handling the social updates.
 Given a set of updated user profiles, our algorithm performs the maintenance mainly in three
 steps:
 (1) update the user profile representation and locate the extended signature-trees of each user by
 hash mapping over the category-entity pairs in his long-term interest list
 (lines \ref{stp-locate-tree-start}-\ref{stp-locate-tree-end});
 (2) update
 the hash table if necessary (line~\ref{stp-hash-update});
 (3) find current user profile from the identified extended signature-trees (line~\ref{stp-find-le}), update the extended signature-tree containing
the current user
profile (lines \ref{stp-update-prof1}--\ref{stp-update-prof2}).
We search the chained hash table, and find the
category-entity pairs that
match those in an updated user profile. If a category-entity pair from the current user profile
can not be found from the hash table, it means a new entity has come and needs to be inserted into
the hash index. The signatures of the tree containing the current user are expanded to include the
unseen entity. To fit the unseen entities, following the classic technique for memory management in
database systems, we reserve $20\%$ space of each entry, and fill it with zones. Hence, we just
increase counters for all entries, until no updating is required. For the leaf entry of the current
user profile, we execute two different
update operations, depending whether the short-term window in the entry is full or not. If the short-term window is not full, we only keep the newly arrived social item in the window. Otherwise, we computes all frequency counters of items in the window, write items in the window into the user profile record, and put the new items in the window. If a user profile can be found from an extended signature-tree, it is an existing user with new activities, and its signature is reconstructed by counting the frequencies of its entities and that of the producers in his browsing list. All the signatures of its ancestor entries are reconstructed based on its new signature. If a user profile does not appear in any signature-tree, it is a new user. We find its block and further to the signature-tree for it to be inserted. As such, the user profiles are well maintained to reflect their recent social activity patterns.


\section{Experimental Evaluation}\label{sec:experiment}
This section evaluates the high effectiveness and efficiency of our proposed social stream recommendation.

\subsection{Experimental Setup}\label{subsec-exp-setup}
We conduct the experiments on four datasets: (1) A real-world dataset {\youtube} that is constructed by crawling the YouTube website using the 20 most popular queries~\cite{DBLP:journals/vldb/ZhouCZQCHW17}. {\youtube} consists of the media data of 787,010 videos that were uploaded to YouTube from 2012 to 2016.
For each video, we crawled its title, description, uploader and its interacted  user information in the ranked results. Producers and consumers for videos are identified according to Definition~\ref{def-user-modes}. (2) A real-world {\dataset{MovieLens} dataset, {\movielen}~\cite{harper2016movielens},
which is publicly available and consists of 20 million user-movie interactions between 138{,}493 users and 27{,}278 movies from  09/01/1995 to 31/03/2015. Since there are no categories or producers available in {\movielen}, we generate them based on our observation on {\youtube} dataset that producers often create social items of one category. We generate the category information by clustering all {\movielen} movies based on their ratings, and regard the users who create social items for one category only and have frequent interactions as producers. (3) A synthetic dataset {\syoutube} created using {\em synthpop}~\cite{synthpop-16} based on {\youtube}. (4) A synthetic set {\smovielen} created with {\em synthpop}~\cite{synthpop-16} based on {\movielen}. The details of these datasets are shown in Table~\ref{tbl-datasets}, including the numbers of producers $|U^p|$,  consumers $|U^c|$, entities $|E|$, interactions $|\mbox{IRact}|$ and social items $|V|$.

\begin{table}[t]
	\centering
	\caption{Overview of datasets. }\label{tbl-datasets}
\begin{tabular}{ccccccc}
\toprule
Dataset &	$|U^p|$ & $|U^c|$ & $|E|$& $C$ &$|\mbox{IRact}|$ & $|V|$\\
\midrule
{\youtubec} & 3,146 & 8.41M & 54,327 & 19 & 49M & 787,010
\\[0.4ex]
{\syoutubec} & 3,146 & 8.41M & 54,327 & 19 & 52M & 787,010
\\[0.4ex]
{\movielenc} & 586 & 138,221 & 28,195 & 15 & 20M & 27,278
\\[0.4ex]
{\smovielenc} & 593 & 138,198 & 28,195 & 15 & 21M & 27,278 \\
\bottomrule
\end{tabular}
\end{table}

\subsection{Evaluation Methodology}
We evaluate our proposed ranking method social stream and user stream Recommendation (ssRec)
in terms of effectiveness and efficiency.
First, we evaluate the effect of the short-term interest window in terms of both
window size $|\mathcal{W}|$  and  the balance parameter
$\lambda_s$ in Equation~\ref{eqn-rec-score-combined}.
We show their sensitivity and the optimal values of the two parameters.
Then we evaluate the impact of using expansion techniques in the recommendation, and that of user profile updates.
Finally, the effectiveness and efficiency of our recommendation approach are evaluated using the optimal parameter settings.

We follow {\citet{wang2018neural}} to set up the stream simulation environment.
We first order all interactions by timestamps, and then evenly split them into six partitions, the first
two of which are the training sets while the other four are reserved for testing.
When the current partition is used for training, its immediate next partition is used for testing.
All effectiveness values are reported when the partition is used for testing only.

We compare our ssRec to  two state-of-the-art baselines,  CTT \cite{huang2016real}
and  UCD~\cite{zanitti2018user}.
CTT fuses collaborative filtering, type and temporal factor together to generate
 recommendation over streams.
UCD is a diversity-based method, where user profiles are expanded with their neighbours.

The effectiveness of all methods in the experiments are evaluated using  precision at  $k$
(\Pat{k})~\cite{wang2018neural} unless specified, which is computed as:
$\Pat{k}= \frac{\# \var{Hit}}{|V|\cdot k}$,
where $k$ is the cutoff in the ranked user list,  \#{\em Hit} is the number of correct
recommendation, and $|V|$ is the number of social items in test data partitions.
We evaluate the efficiency of different approaches based on the average response time for an item on the stream. All experiments are conducted on a server using an Intel Xeon E5 CPU with 25 GB RAM
running RHEL v6.3 Linux.
We use TagMe~\cite{tagme} for entity extraction and implement the
recommendation process over Apache Storm.

\subsection{Effectiveness Evaluation}
We first compare our BiHMM model with traditional HMM model to verify the dependency of user interests on the media producers. Then, we evaluate the effect of parameters, $|\mathcal{W}|$ and $\lambda_s$, by conducting the $R$-based recommendation to find their optimal values. Finally, we compare our proposed approach with the state-of-art stream recommendation approaches, and evaluate the effect of user profile updates in our approach.

\subsubsection{Comparing BiHMM and HMM}
We prove the superiority of our proposed BiHMM by comparing with HMM.
The number of optimal hidden states are tuned based on the user browsing history. We divide the browsing history of users based on their media browsing time into two parts: the first 80\% historical data in the profile for training and the latter 20\% history data for testing.
Note that, here we consider users' interaction information in both producer  and consumer modes. Our model is evaluated by {\em Accuracy}, which is the correct prediction percentage of a user's next interest category among all.
For each user, we decide the optimal number of hidden states over HMM by testing the {\em Accuracy} of the model at different state number values from 1 to a number where the {\em Accuracy} reaches the peak. The optimal number of hidden states is obtained when the highest {\em Accuracy} is achieved. Using the optimal hidden state number of each consumer user, we train our BiHMM model by embedding the hidden states of producers appearing in each browsing history, and obtain the optimal parameters for BiHMM, including the initial state probability distribution, the state transition probability matrix and the state transition probability matrix.

\begin{figure}[t]
\centering
\subfigure[{\youtubec}]{\label{fig:a}\includegraphics[width=0.46\linewidth]{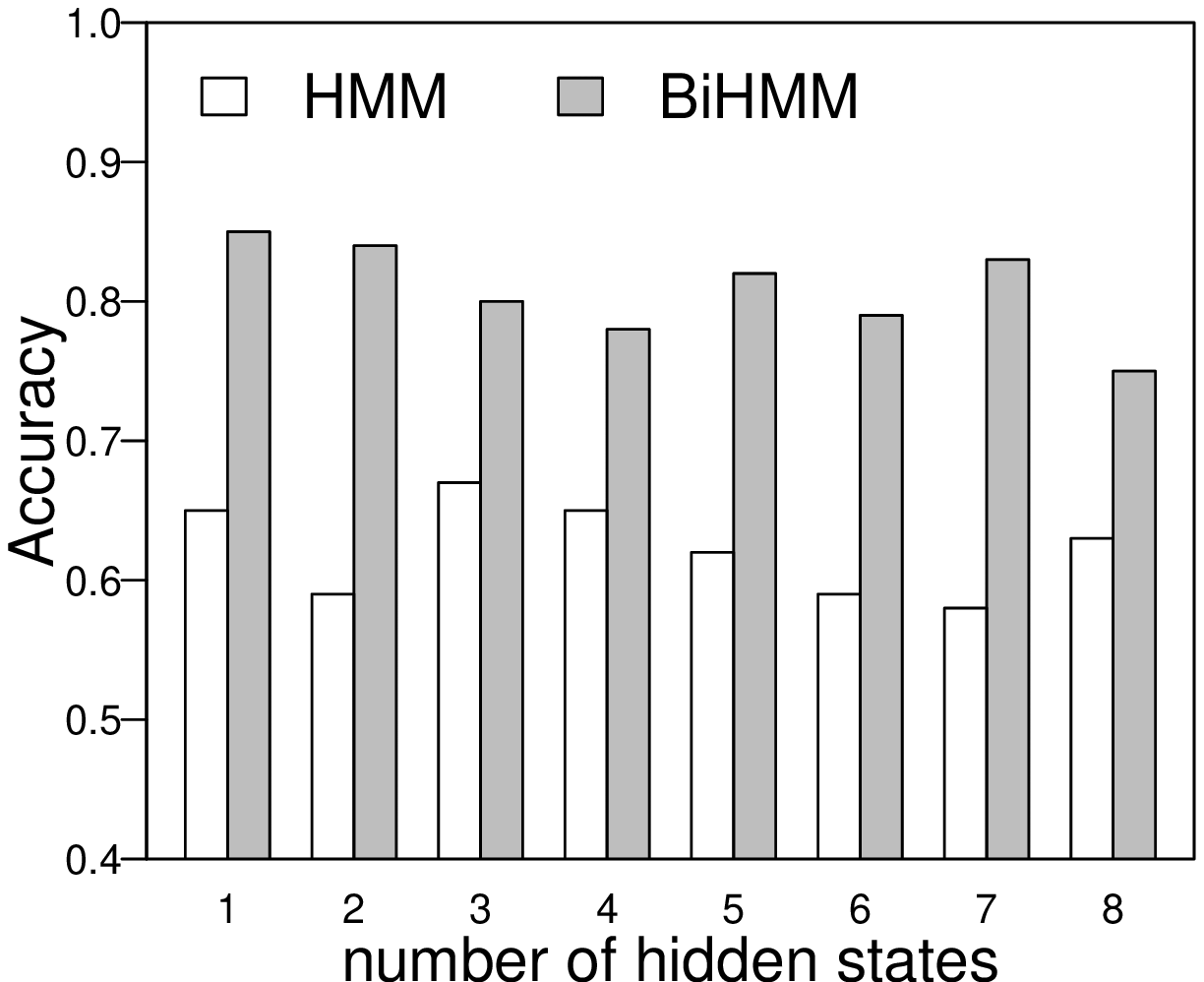}}
\subfigure[{\syoutubec}]{\label{fig:b}\includegraphics[width=0.46\linewidth]{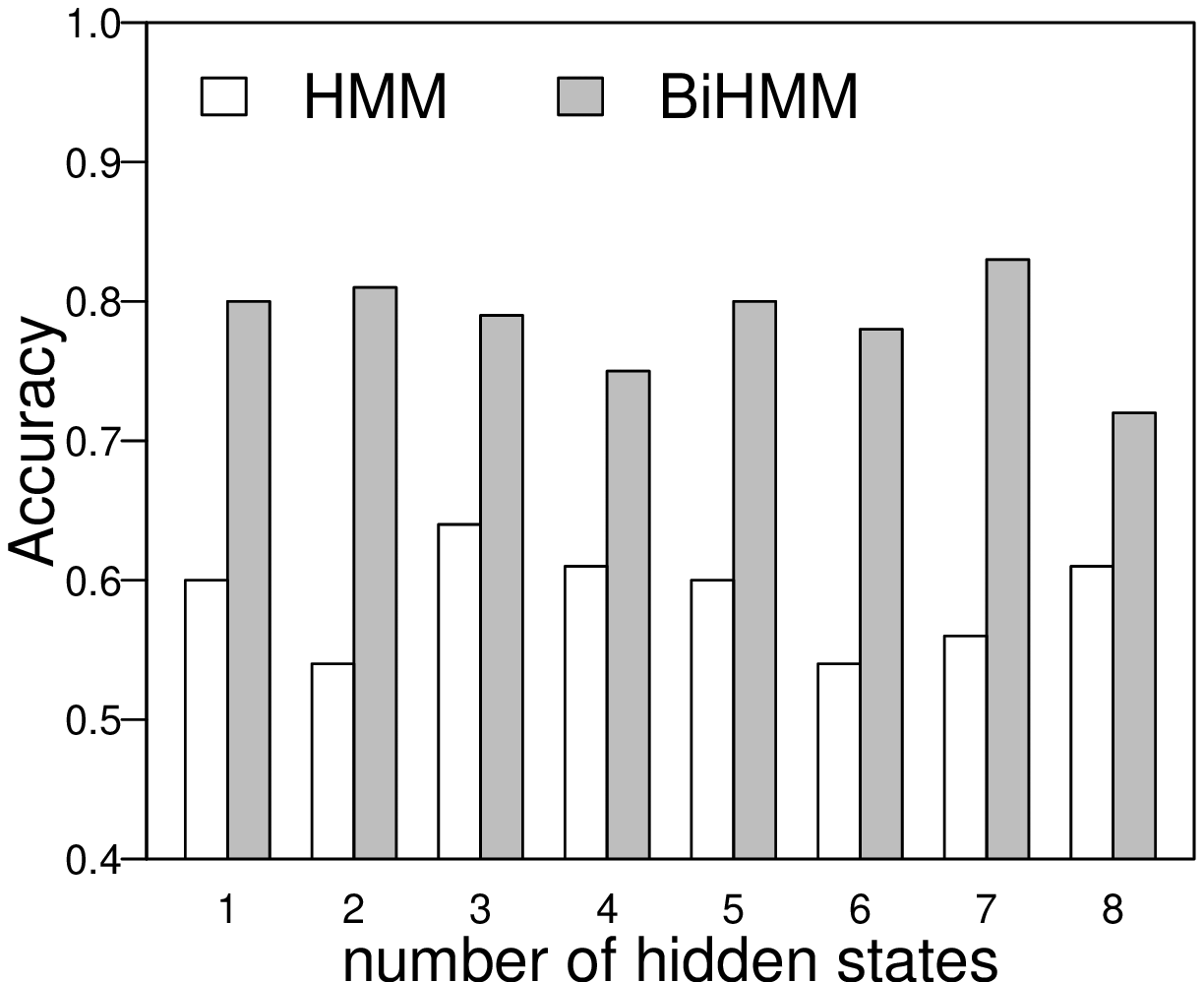}}
\subfigure[{\movielenc}]{\label{fig:b}\includegraphics[width=0.46\linewidth]{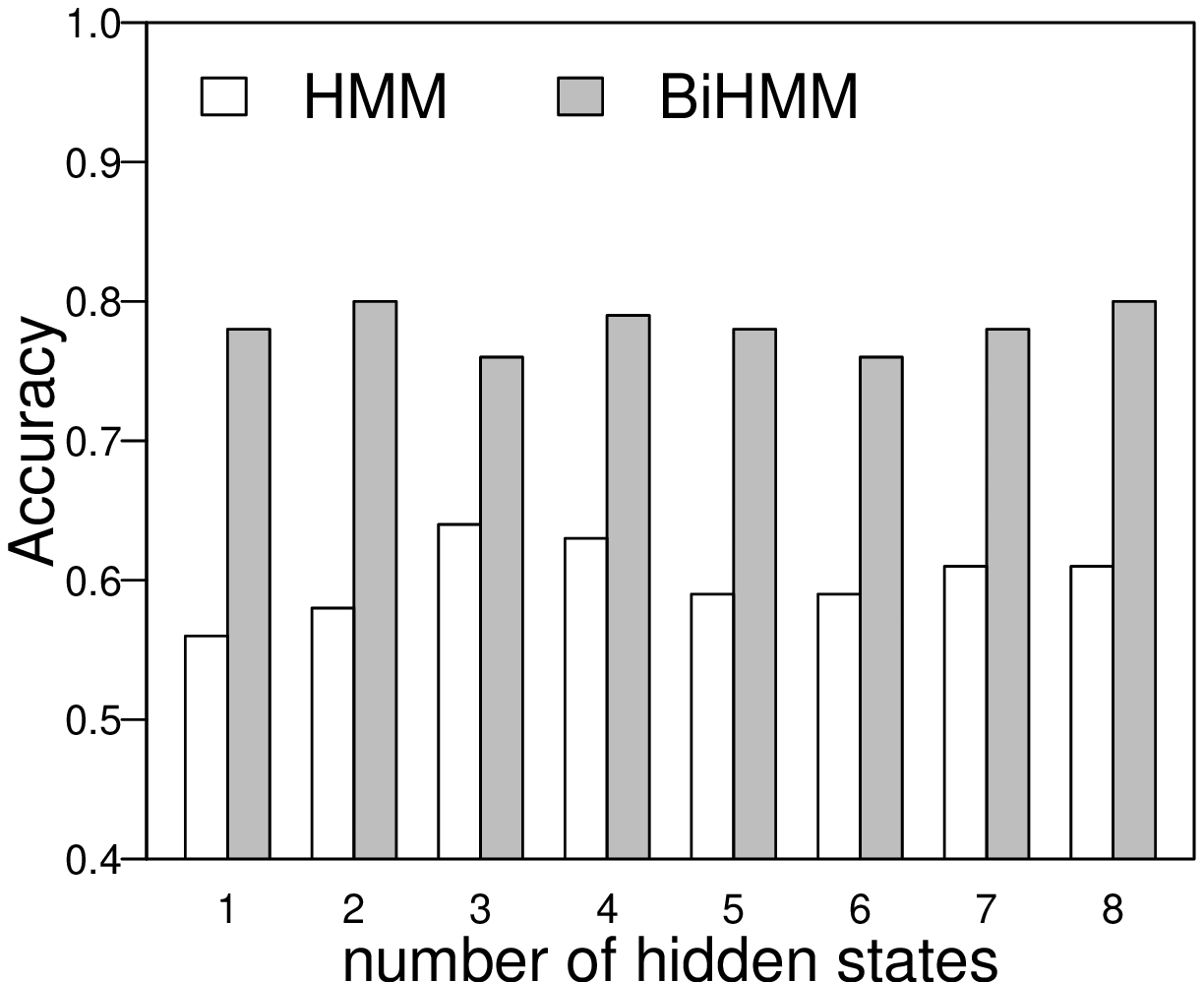}}
\subfigure[{\smovielenc}]{\label{fig:b}\includegraphics[width=0.46\linewidth]{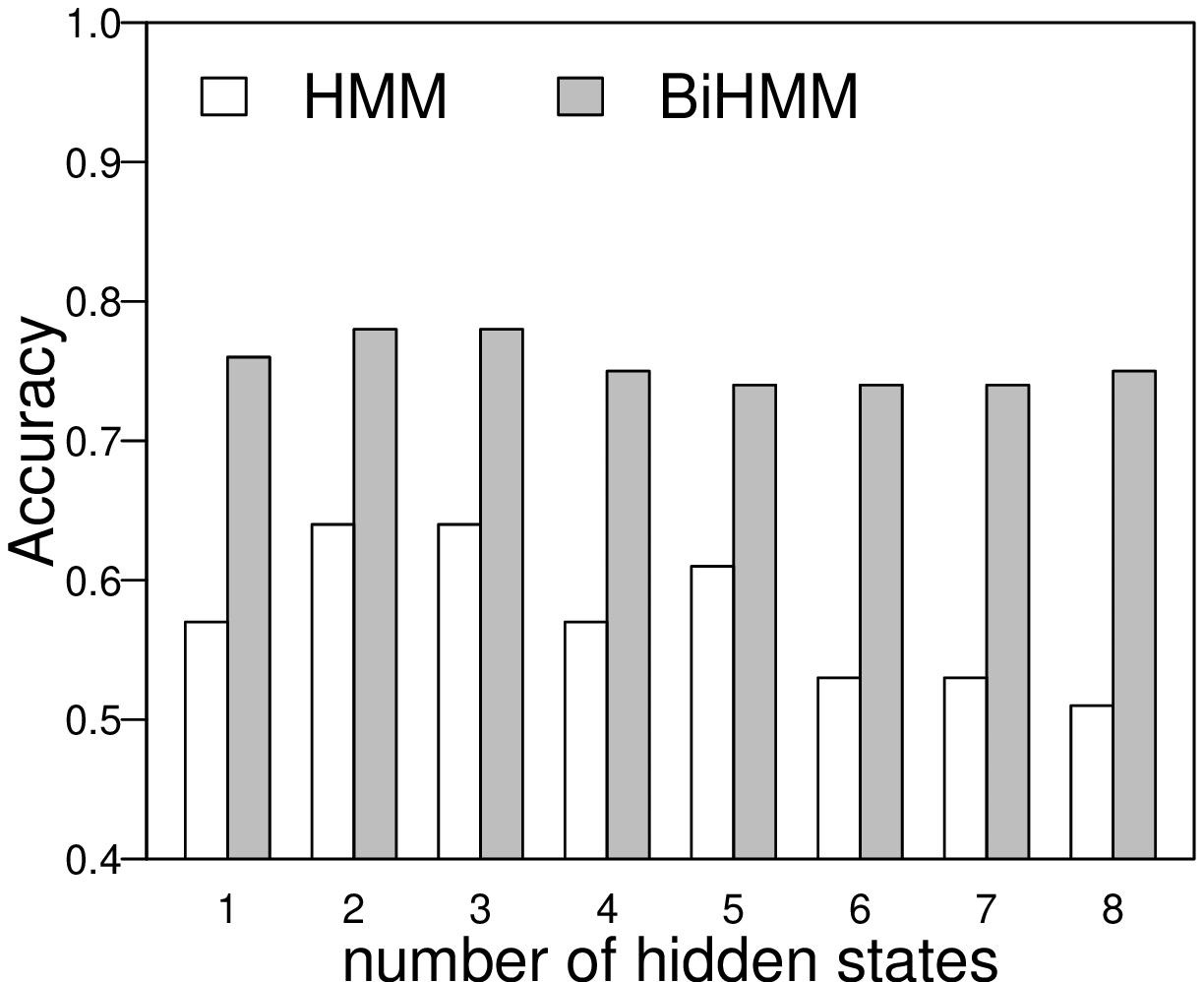}}\vspace{-1ex}
\caption{Effectiveness comparison between  BiHMM and HMM.}
\label{fig:bihmm}\vspace{0ex}
\end{figure}

We test the prediction accuracy of BiHMM and HMM for each consumer, put the users with the same optimal hidden state number into the same group, and report the prediction results for different groups. Fig.~\ref{fig:bihmm} shows the prediction results of two models for groups with optimal hidden state numbers from 1 to 8. From the figure, we can observe the same trend across four datasets and for all
users -- that the BiHMM is better than the HMM. The results have further verified our hypothesis that consumers' interests are dependent on the producers as well, which is not captured by HMM.

\subsubsection{Effect of $|\mathcal{W}|$}
We evaluate the impact of the short-interest window size $|\mathcal{W}|$ over our simulated stream data to find the optimal $|\mathcal{W}|$. We test the prediction precision ({\Pat{k}}) of the recommendation by varying $|\mathcal{W}|$ from 1 to 10, where $|\mathcal{W}|$ means the number of recently browsed items in a window. At each $|\mathcal{W}|$ value, we measure the prediction precision of recommendation by changing the weight of short-term interest measure $\lambda_s$ from 0.1 to 1 with step 0.1, and report the optimal precision value. The prediction precisions for one partition are calculated based on whether the recommendation is accepted by the users in its next partition.
For example, if we tune $|\mathcal{W}|$ on the first partition, then we evaluate the prediction precision using the data in the second partition and
only keep the {\em \#Hit} in it until we complete the tests over all partitions. After all partitions are used in the test, we compute {\Pat{k}} by considering all testing partitions, and the results are reported in Fig.~\ref{fig:w}.

\begin{figure}[t]
\centering
\subfigure[{\youtubec}]{\label{fig:a}\includegraphics[width=0.46\linewidth]{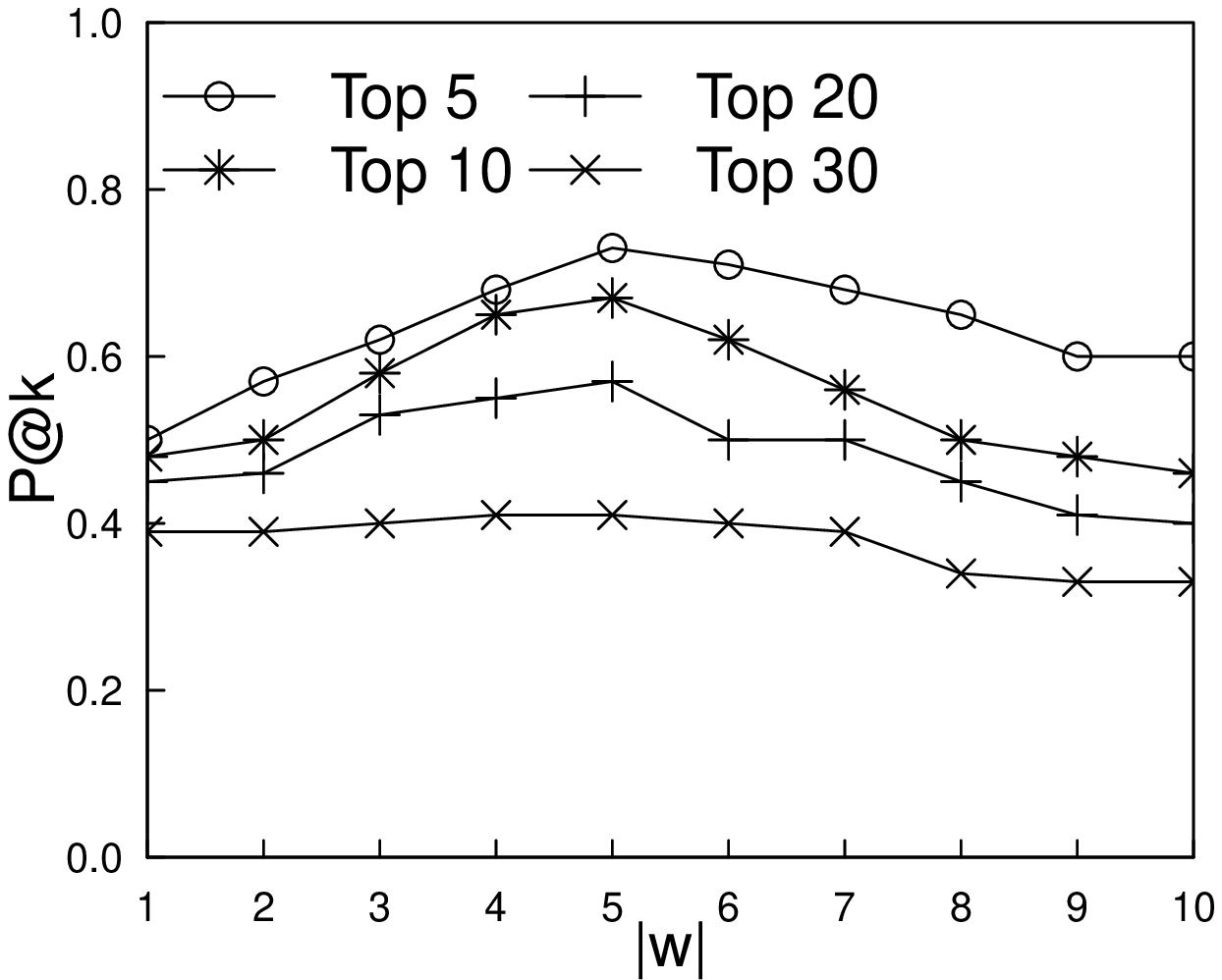}}
\subfigure[\syoutubec]{\label{fig:b}\includegraphics[width=0.46\linewidth]{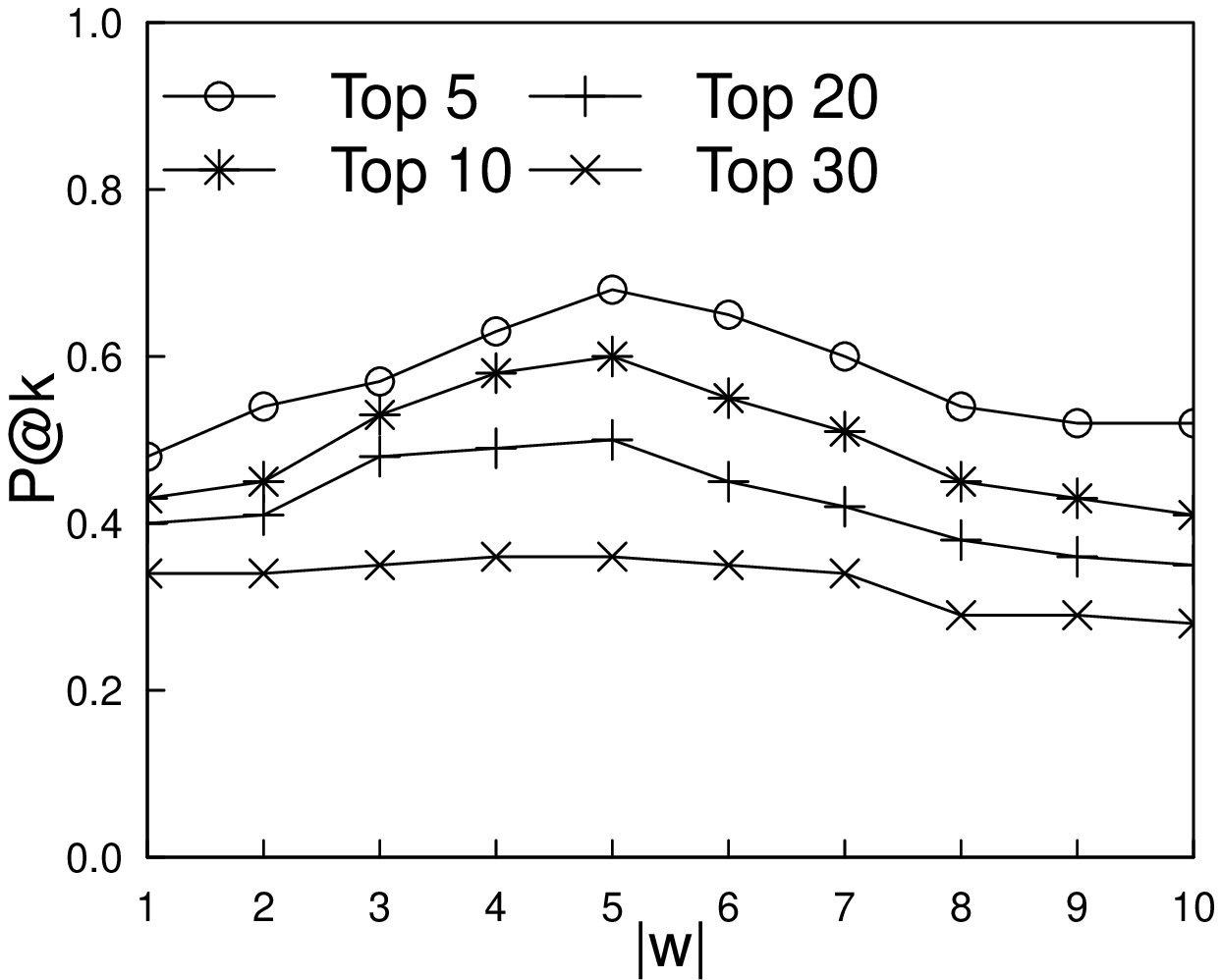}}
\subfigure[\movielenc]{\label{fig:b}\includegraphics[width=0.46\linewidth]{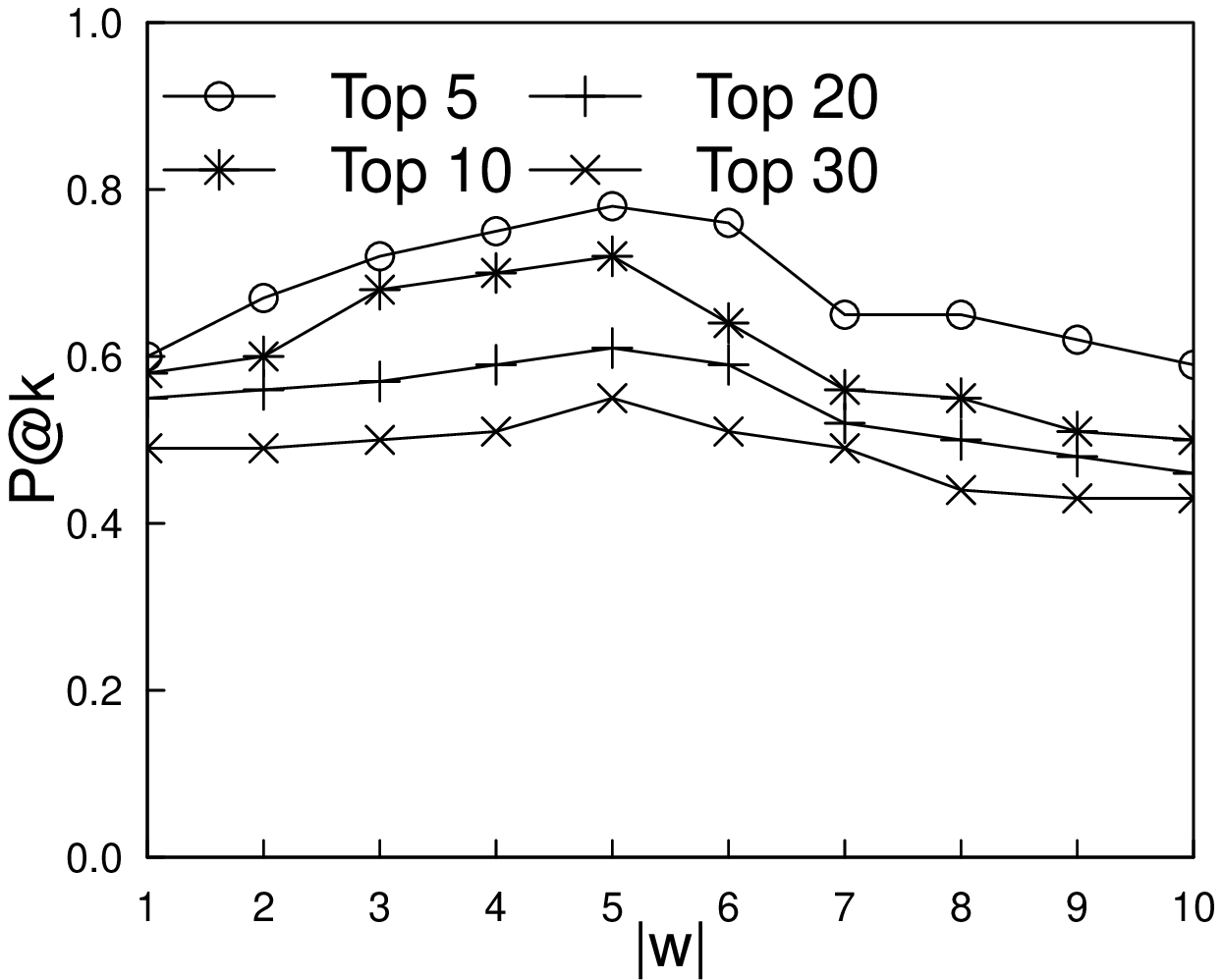}}
\subfigure[\smovielenc]{\label{fig:b}\includegraphics[width=0.46\linewidth]{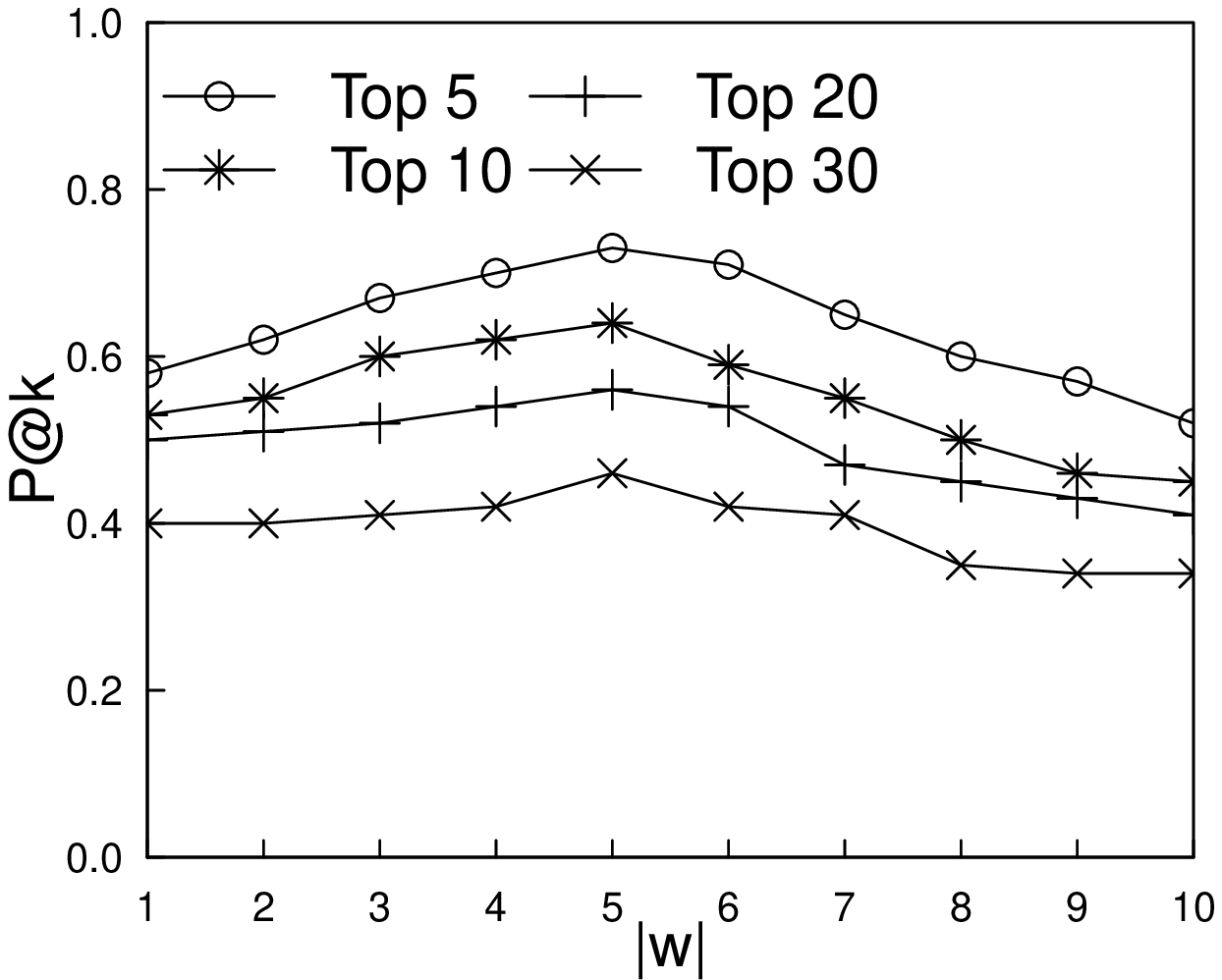}}\vspace{-1.5ex}
\caption{Effect  of short-term interest window size}
\label{fig:w}
\end{figure}
Clearly, when a small $|\mathcal{W}|$ is adopted, the user short-term interests are not accurately predicted due to the interest drift. On the other hand, if a large $|\mathcal{W}|$ is employed, the short-term  interest may fall back to the long-term interest. The optimal effectiveness is always achieved when $|\mathcal{W}|=5$. Thus we set the default $|\mathcal{W}|$ to 5 and use it in all following tests.

\subsubsection{Effect of $\lambda_s$}
\begin{figure}[t]
\centering
\subfigure[{\youtubec}]{\label{fig:a}\includegraphics[width=0.46\linewidth]{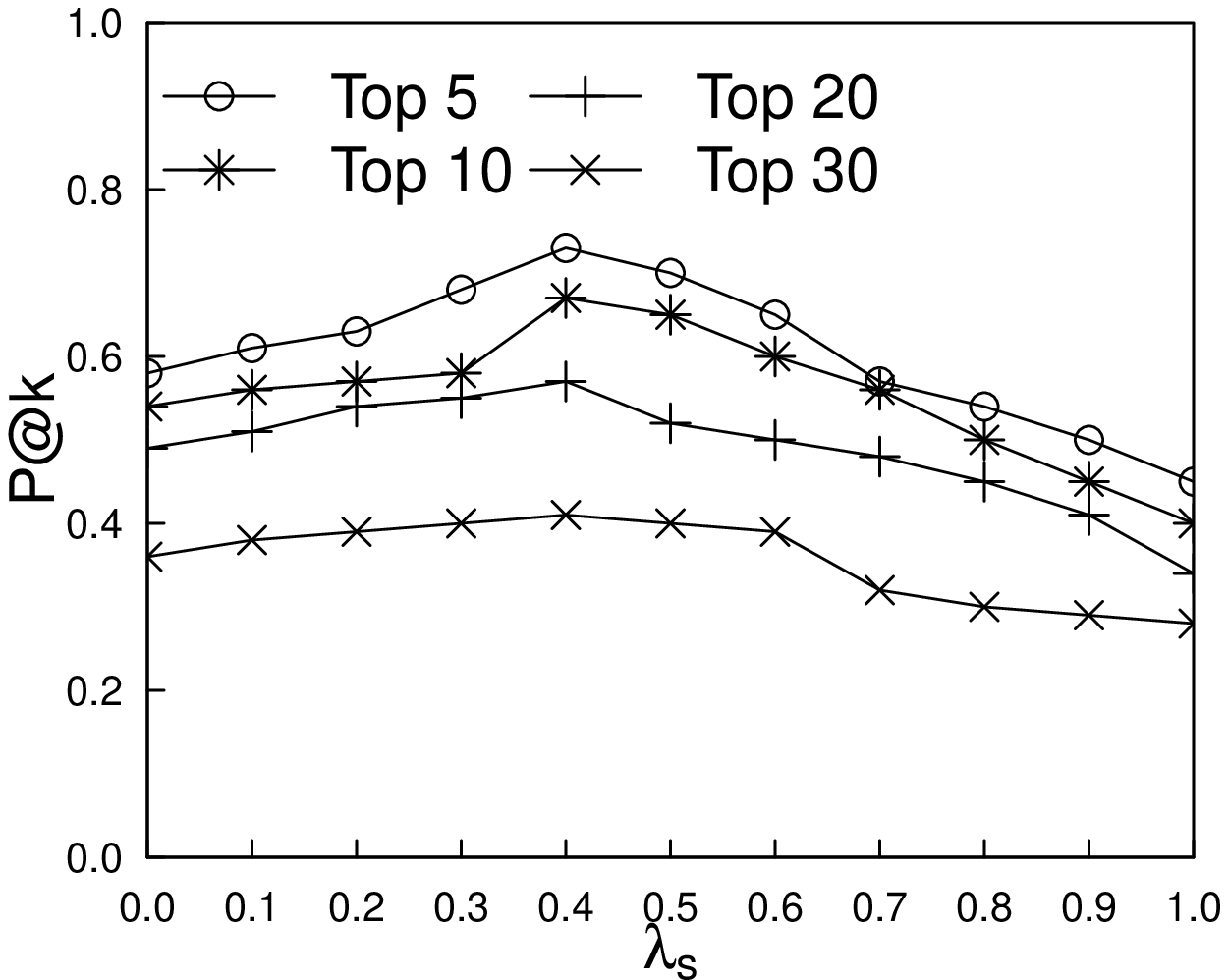}}
\subfigure[{\syoutubec}]{\label{fig:b}\includegraphics[width=0.46\linewidth]{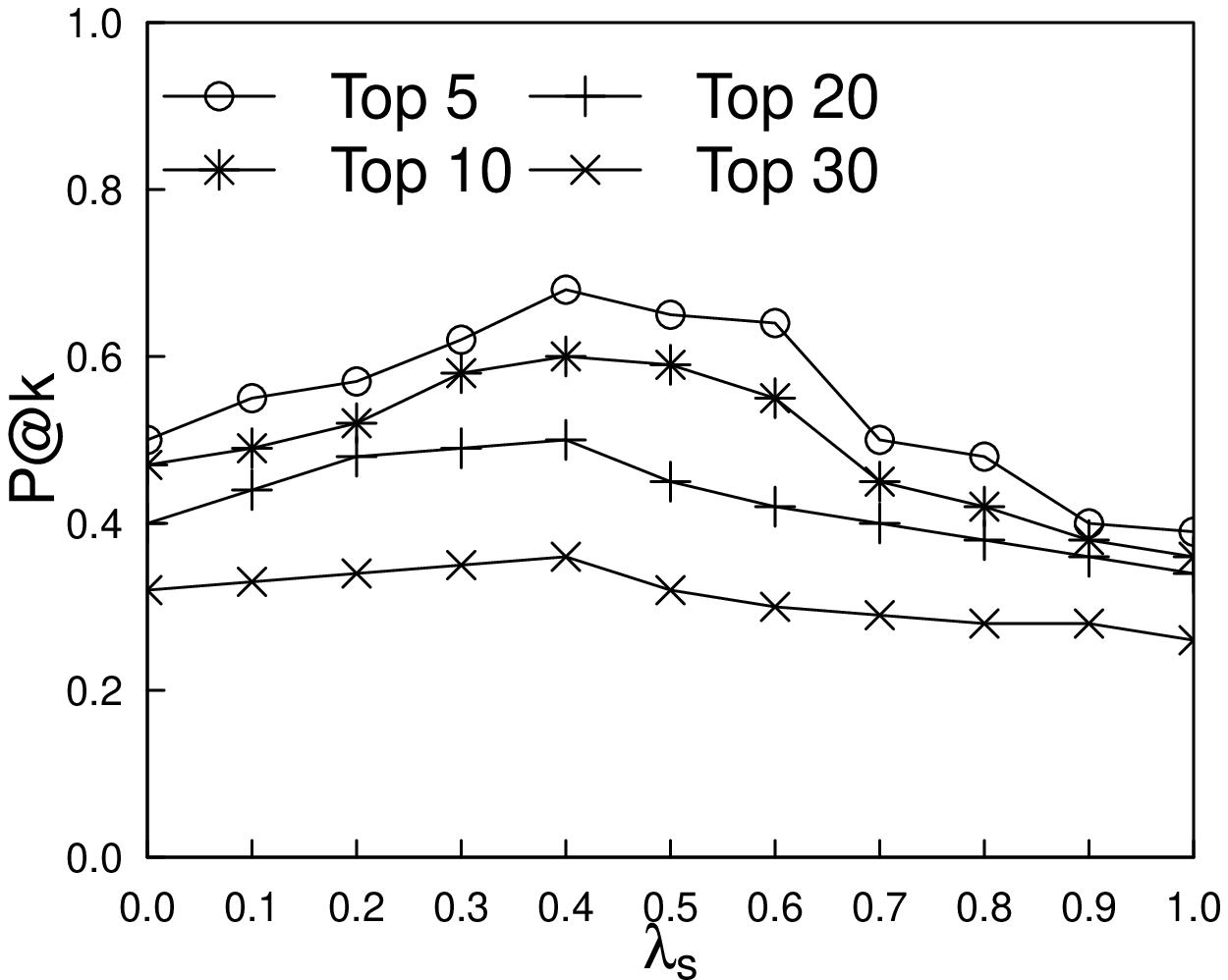}}
\subfigure[{\movielenc}]{\label{fig:b}\includegraphics[width=0.46\linewidth]{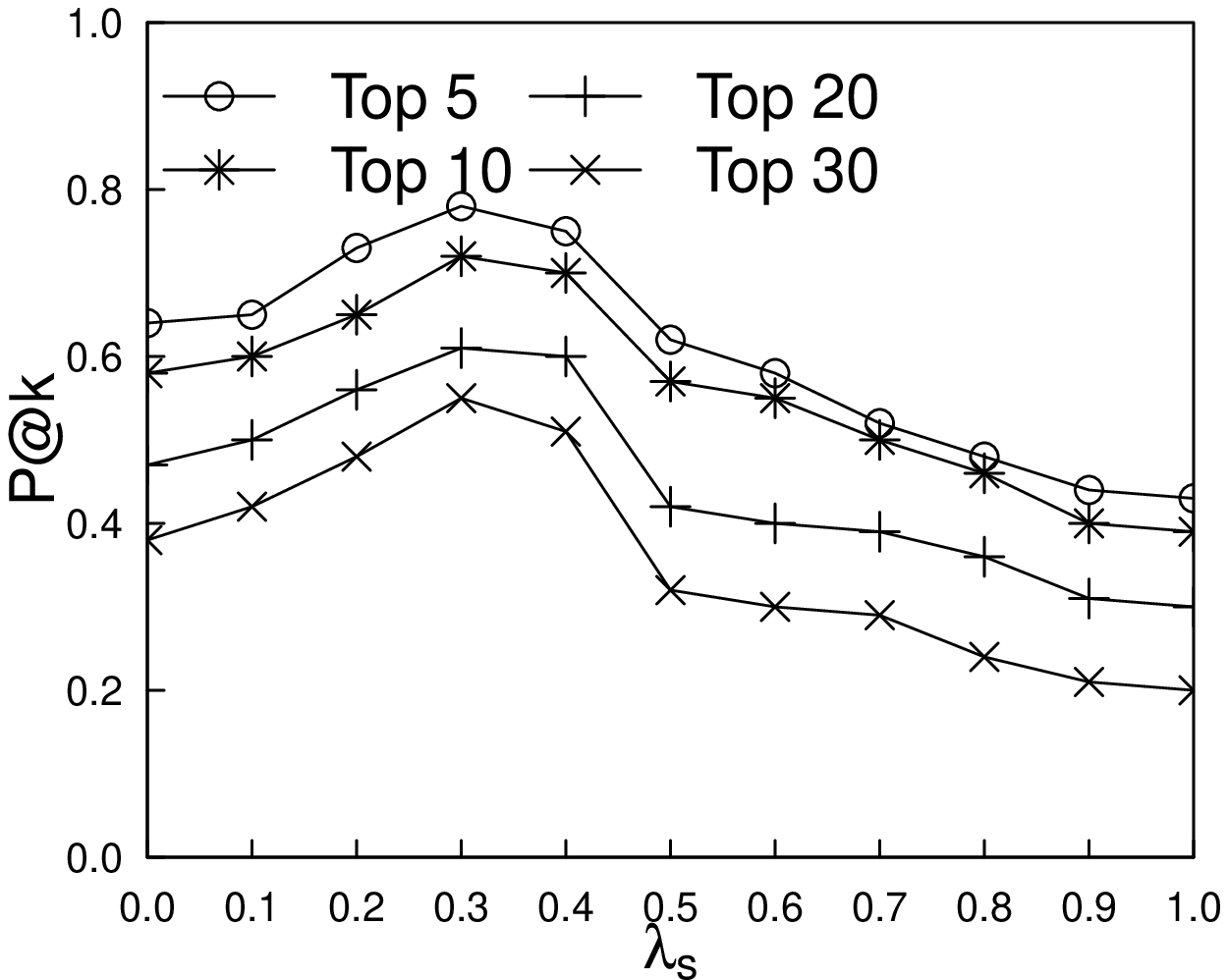}}
\subfigure[{\smovielenc}]{\label{fig:b}\includegraphics[width=0.46\linewidth]{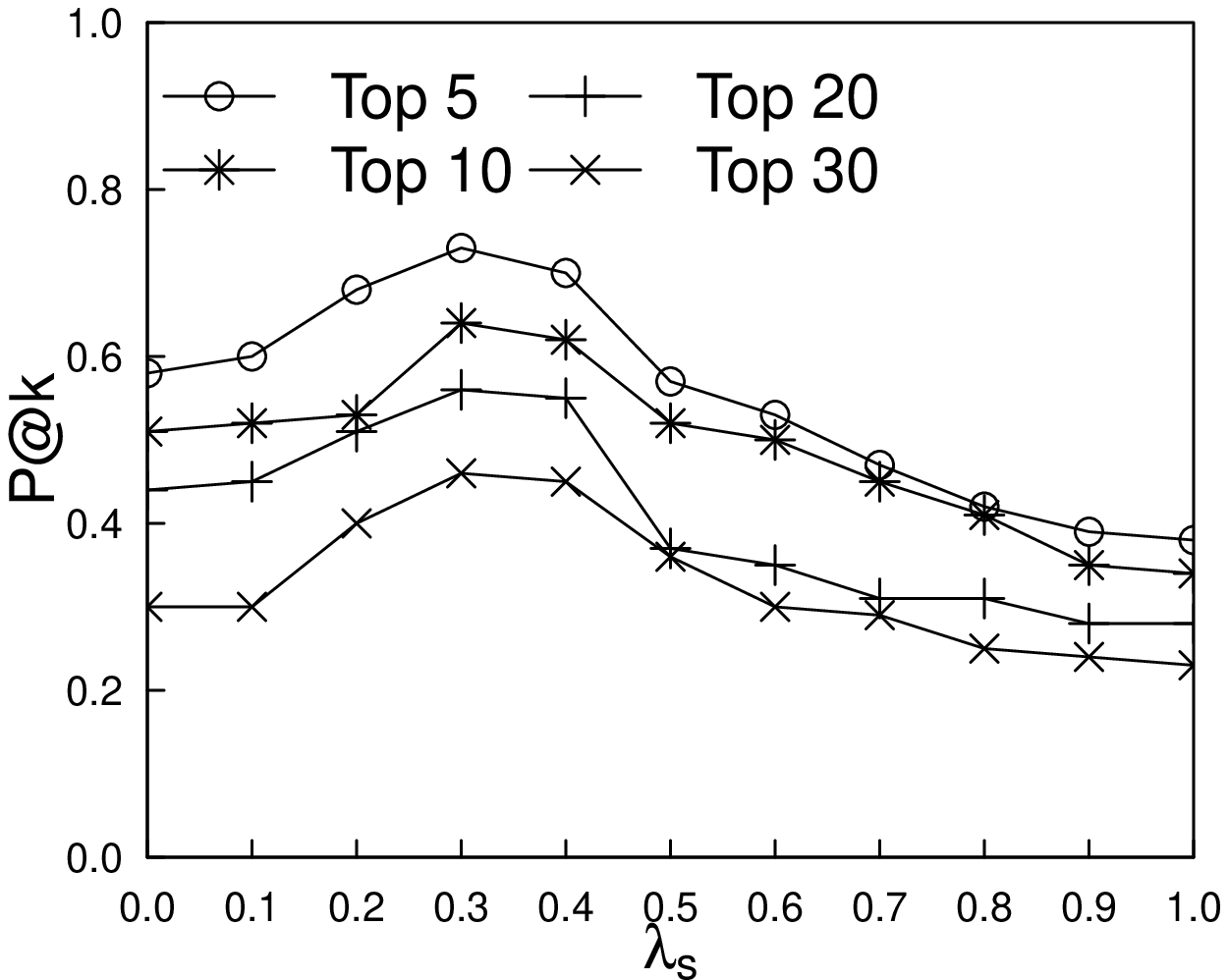}}\vspace{-1ex}
\caption{Effect of short-interest weight $\lambda_s$}
\label{fig:lambda}\vspace{0ex}
\end{figure}
As our final recommendation score consists of both short-term and long-term components, the balance
parameter $\lambda_s$ in Equation~\ref{eqn-rec-score-combined} may be important.
We apply the same parameter tuning settings over simulated item streams as previously discussed,
with the short-term interest window fixed to 5. The test results on the test sets only are reported in
Fig.~\ref{fig:lambda}. As we can observe, the recommendation effectiveness is increased with the increase of $\lambda_s$, reaches an optimal point, and then decreases for each of the four datasets. We obtain the optimal $\lambda_s$ values for two types of datasets, which are 0.4 and 0.3 for {\youtube} and {\movielen} respectively.
Since the two synthetic datasets are generated according to the original data distribution, they
have the same optimal settings as their original dataset.
A larger $\lambda_s$ on {\youtube} also suggests that users' interests are less robust on {\dataset{YouTube}} than on the {\dataset{MovieLens}} website, which is intuitive as items on {\dataset{YouTube}} are also created more quickly than on the {\dataset{MovieLens}}.

\subsubsection{Recommendation Effectiveness Comparison}

%

We use the optimal settings obtained from our previous experiments and evaluate our final recommendation effectiveness by comparing with existing competitors, CTT and UCD.
As described in Section~\ref{sec:streamuser-match}, entity-based expansion is applied to introduce diversity in recommendation.
To better show the effectiveness gain of using the expansion techniques, we report the results of our alternative ssRec-ne that neglects the entity expansion as a reference. We recommend the streaming items to top $k$ users, where $k$ is set to 5, 10, 20 and 30 respectively. Fig. \ref{fig:effectivness} shows the comparison results.

\begin{figure}[t]
\centering
\subfigure[\youtubec]{\label{fig:a}\includegraphics[width=0.46\linewidth]{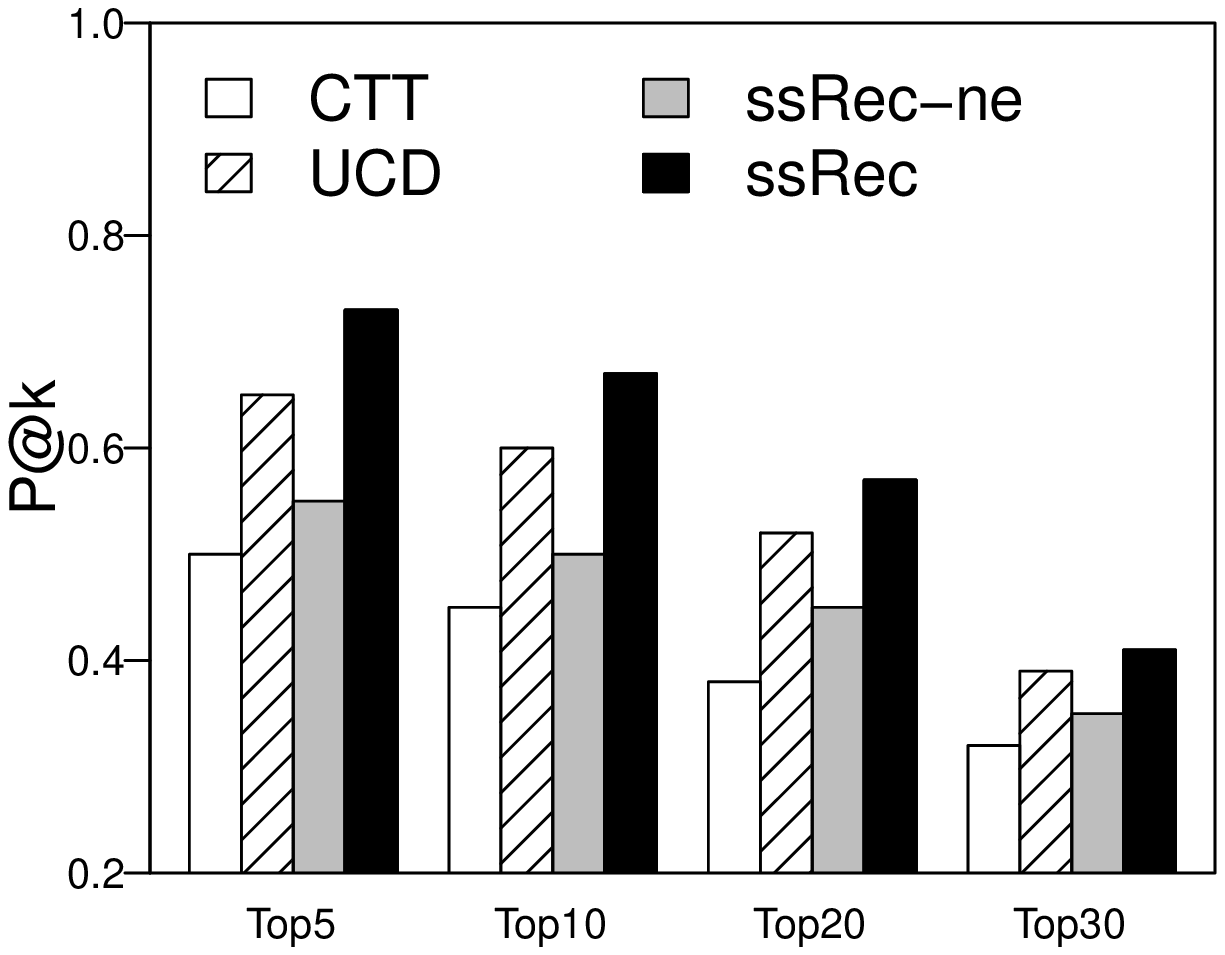}}
\subfigure[\syoutubec]{\label{fig:b}\includegraphics[width=0.46\linewidth]{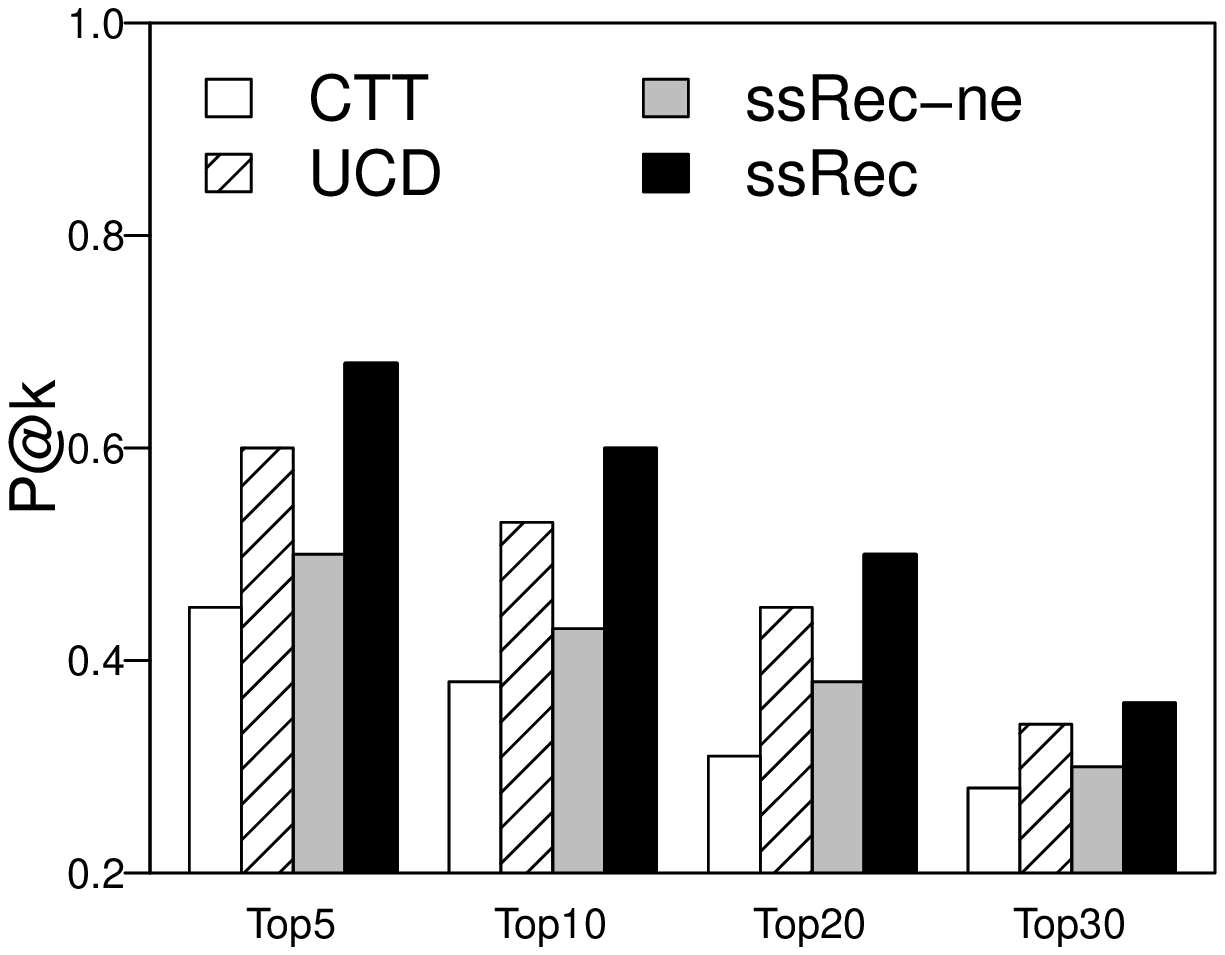}}
\subfigure[\movielenc]{\label{fig:b}\includegraphics[width=0.46\linewidth]{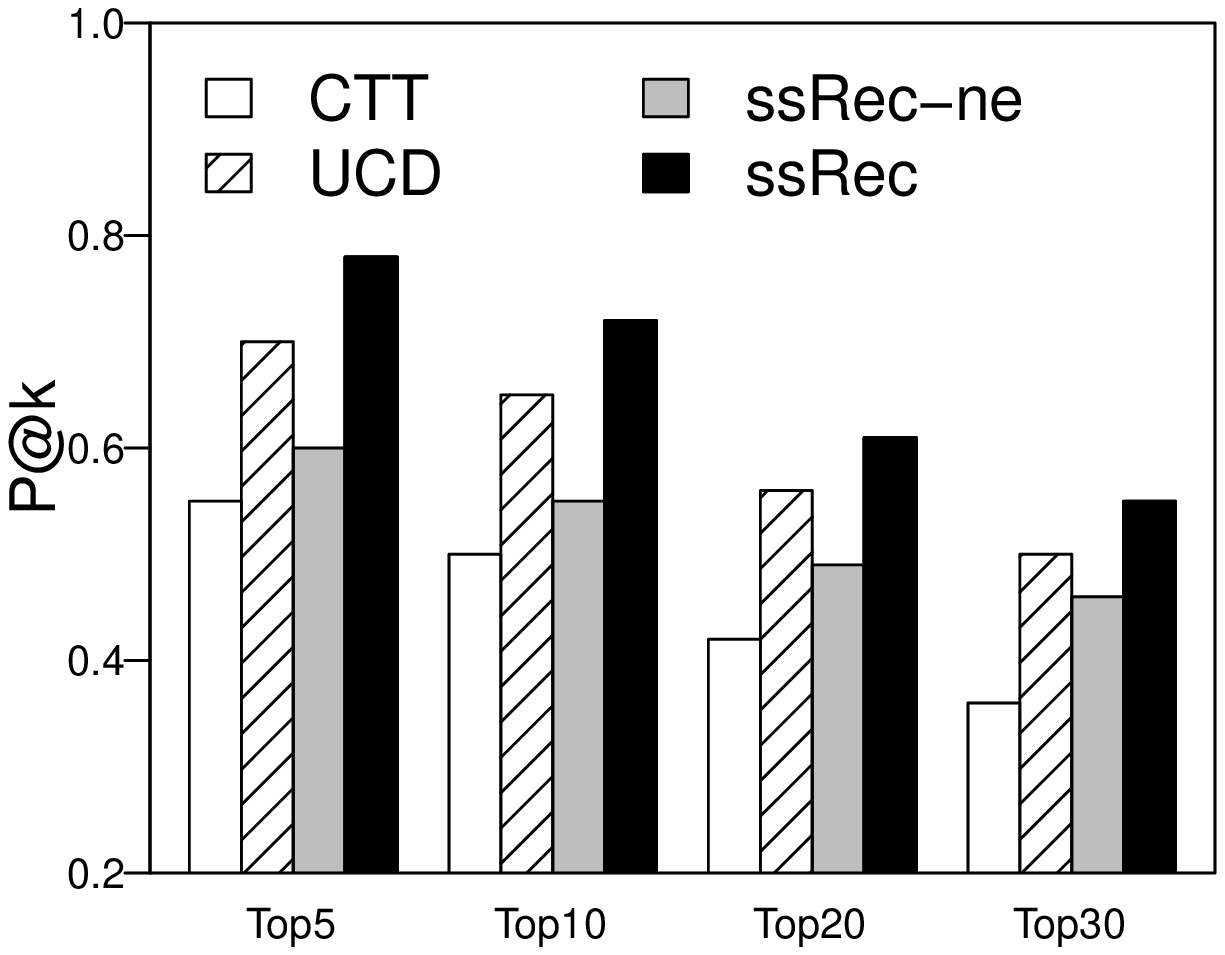}}
\subfigure[\smovielenc]{\label{fig:b}\includegraphics[width=0.46\linewidth]{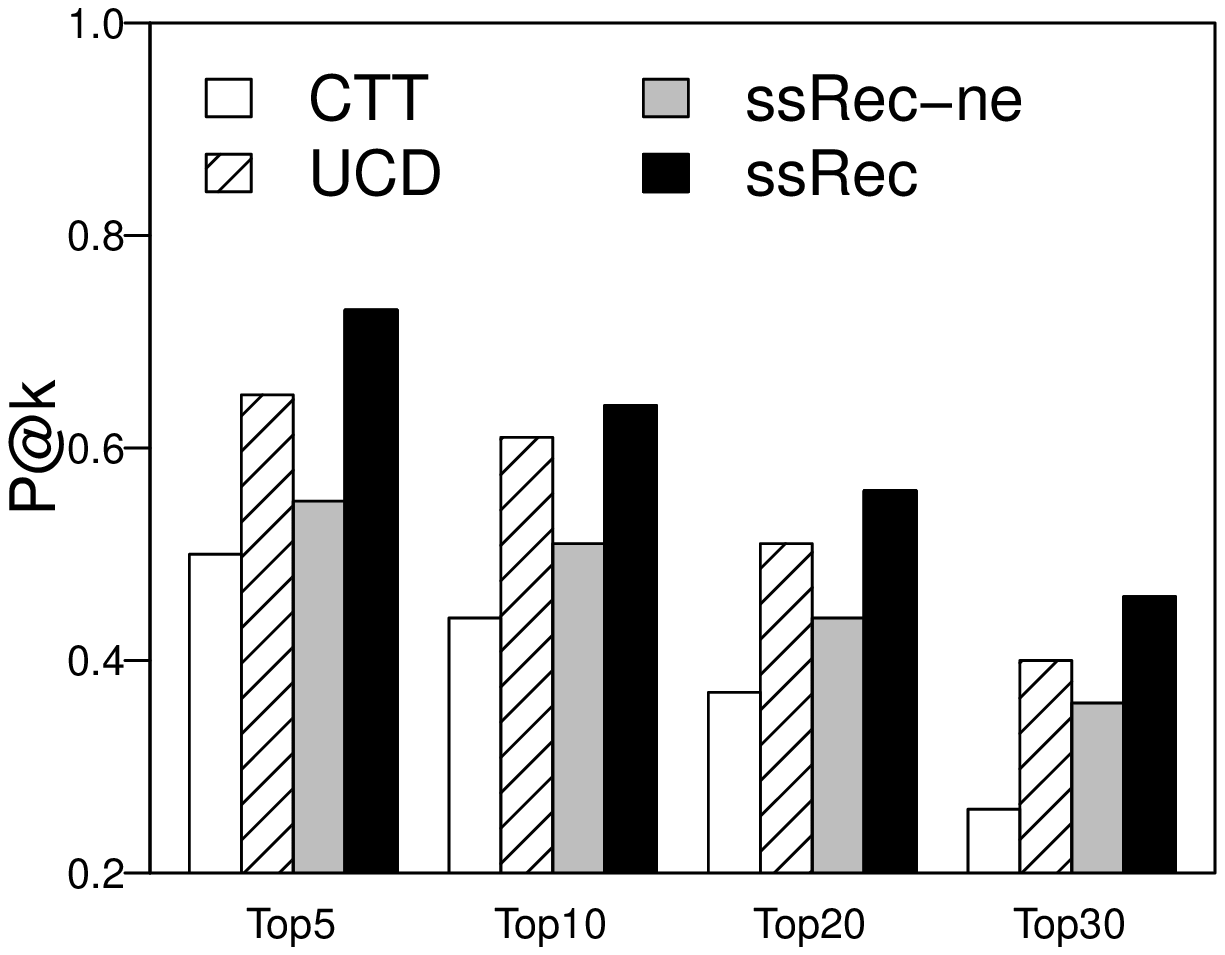}}\vspace{-1.5ex}
\caption{Effectiveness comparison}
\label{fig:effectivness}
\end{figure}

As we can see, our social stream and user steam recommendation approach (ssRec)
achieves a much better performance on all four datasets compared with
the other alternative, the stream recommendation without entity expansion (ssRec). It is because the expansions exploit more entities closely related to user's interests, which reveal user's potential interests. Without entity expansion, the system only recommends the items based on the exact matched entities, which limits user's interest into a narrow scope, resulting in a low recommendation precision. Comparing with existing competitors, our ssRec approach performs best at all $k$ settings among all considered methods, and the improvement is consistent across all four datasets. This is because we consider both the short-term and long-term interests of users in terms of their social properties (producer-consumer dependencies) and item contents (entity expansion) in their interest prediction, which provides a complete representation of user's preferences. CTT performs worst because it ignores the user's short-term interest and the diversity of item-user interaction. Thus, the users' recent interested items cannot be recommended. Meanwhile, ignoring the diversity of recommendation leads to a resulting list containing almost same items, which does not reflect the complete view of user interests.
Although UCD exploits diversity-based user profile to find more diverse items for users, it neglects the significance of short-term interest as what CTT does, leading to a lower $P@k$. All these confirm that our proposed method is superior to other competitors in terms of effectiveness.

\subsubsection{Effect of User Profile Updates}
We test the effect of user profile updates on the effectiveness of recommendation over four test collections. For each collection, we consider two settings: (1) a stream setting on which the model is updated from the previous partition (ssRec); and
(2) a static setting on which the model is trained on the training set and the update operations are ignored (ssRec-nu).
We measure the effectiveness of our recommender system under two settings on the four test partitions. Fig.~\ref{fig:update-effectiveness} shows that the effectiveness changes with respect to different top-$k$ target users. As we can observe, with user profile updates, we obtain a big effectiveness gain on {\Pat{k}}. This is because with the updates in user profile, the user's long-term and short-term interests can be well captured. Without updates, user's profiles do not reflect their recent activity patterns. The improvement of ssRec over ssRec-nu confirms the importance of dynamic maintenance.

\begin{figure}[t]
\centering
\subfigure[\youtubec]{\label{fig:a}\includegraphics[width=0.46\linewidth]{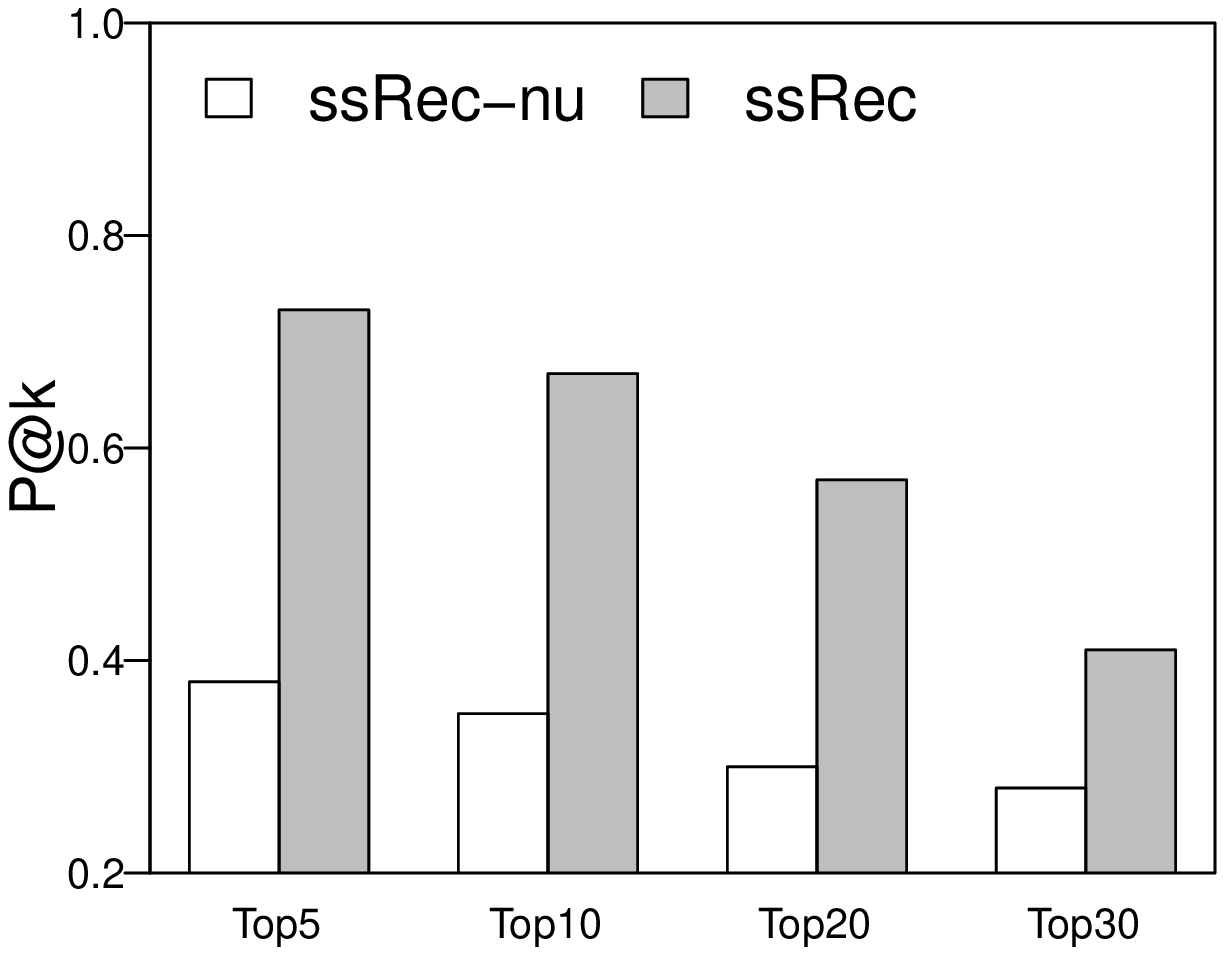}}
\subfigure[\syoutubec]{\label{fig:b}\includegraphics[width=0.46\linewidth]{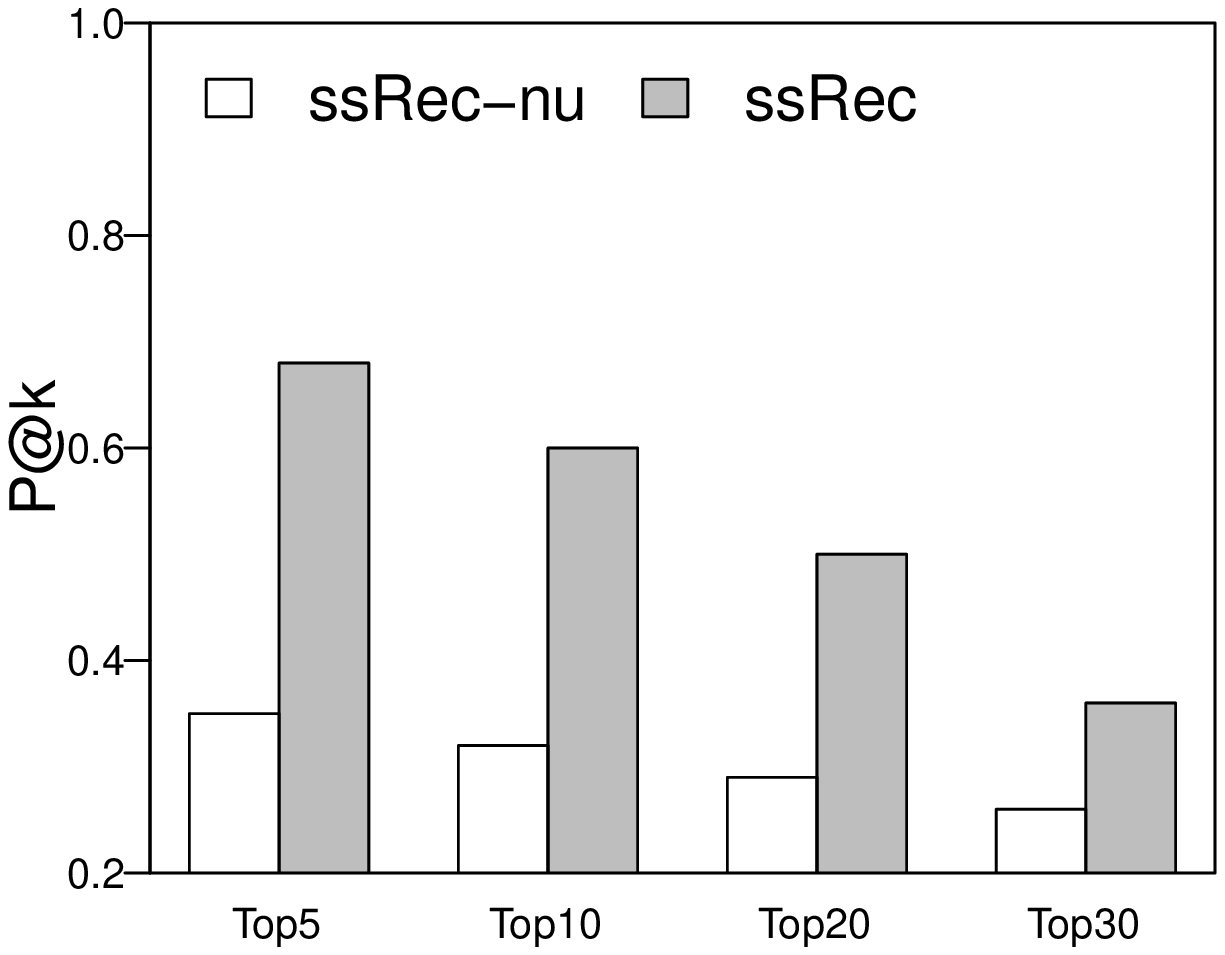}}
\subfigure[\movielenc]{\label{fig:b}\includegraphics[width=0.46\linewidth]{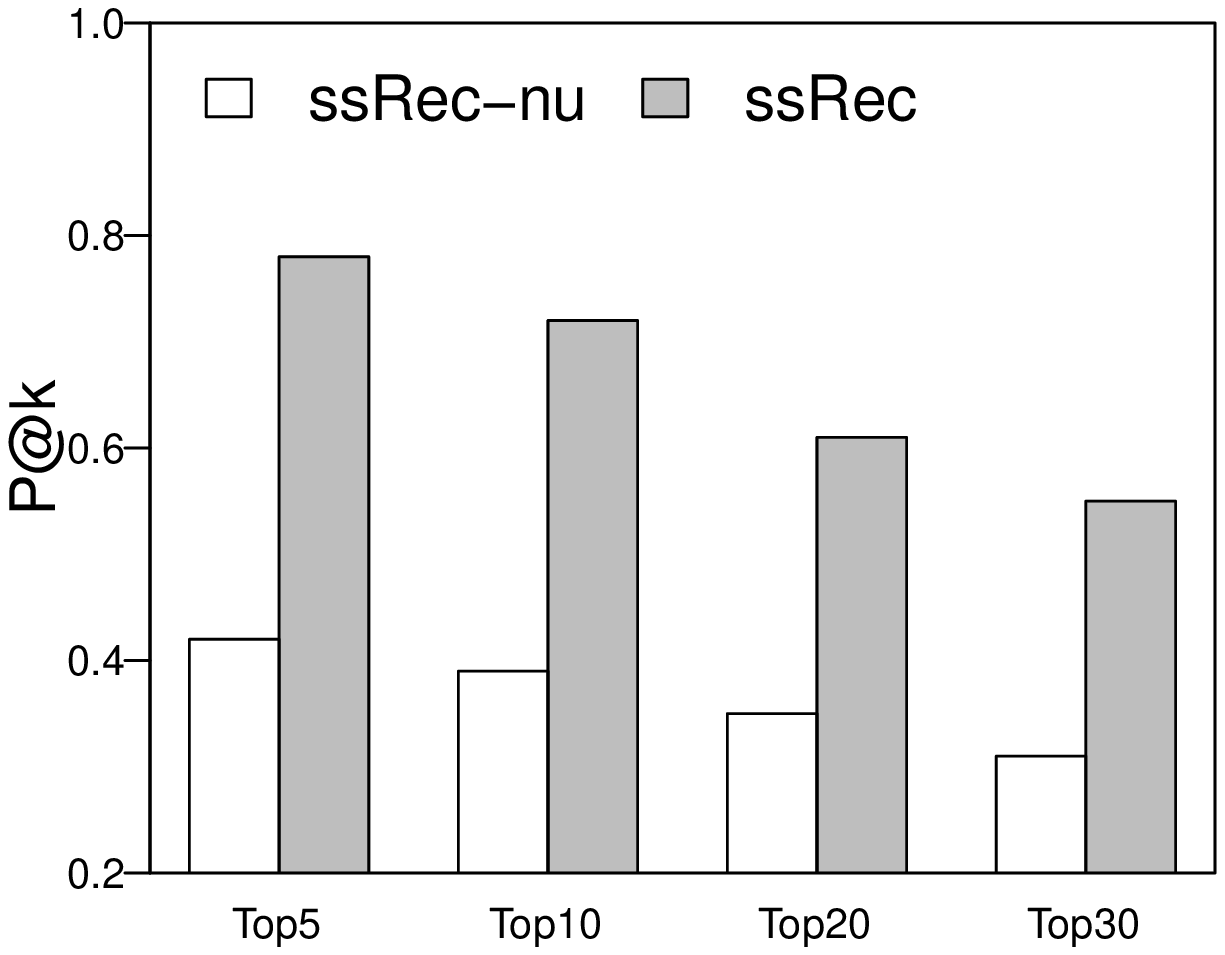}}
\subfigure[\smovielenc]{\label{fig:b}\includegraphics[width=0.46\linewidth]{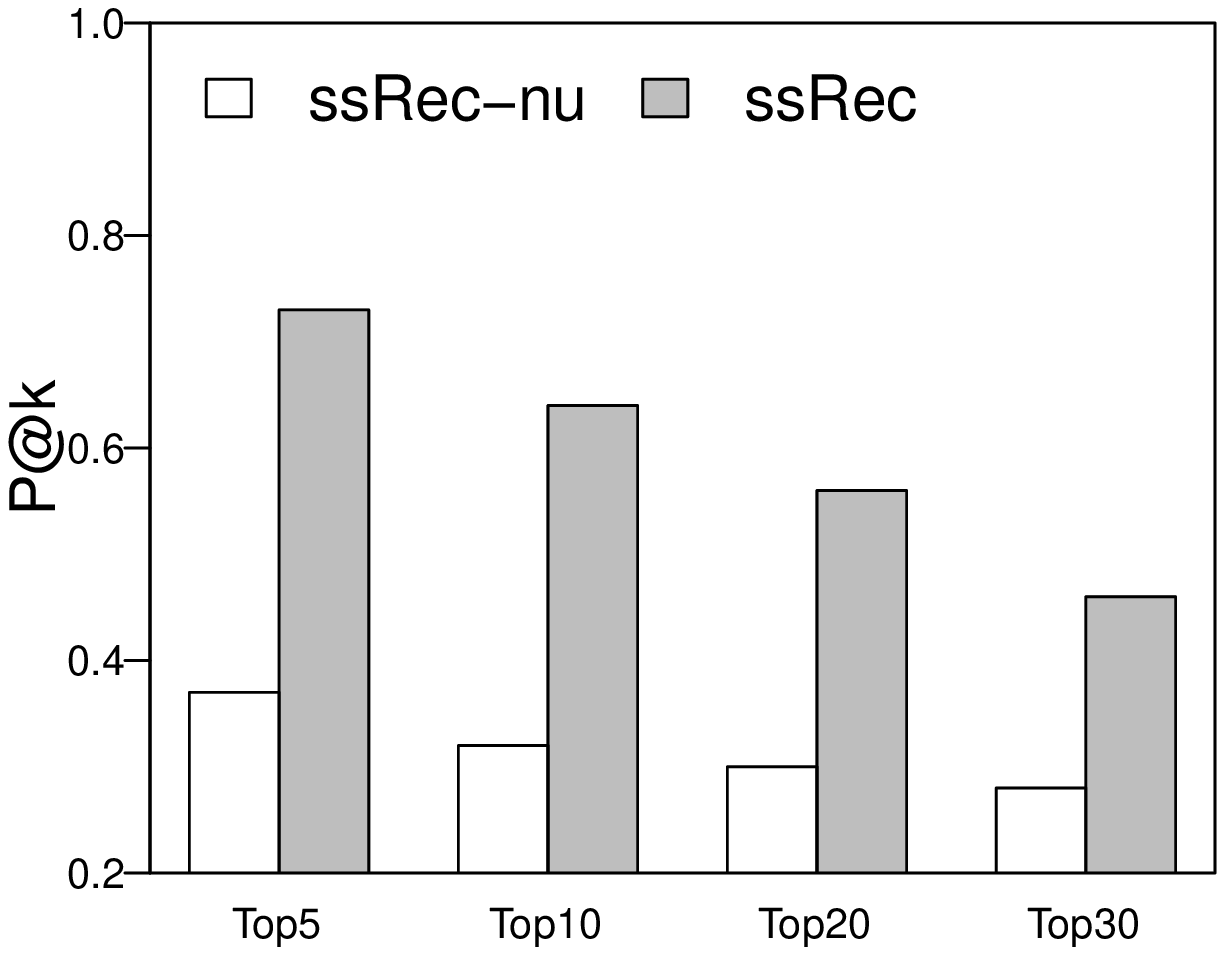}}\vspace{-1.5ex}
\caption{Effect of user profile updates}
\label{fig:update-effectiveness}\vspace{0ex}
\end{figure}

\begin{figure}[h]\vspace{-2ex}
\centering
\subfigure[\youtubec]{\label{fig:a}\includegraphics[width=0.46\linewidth]{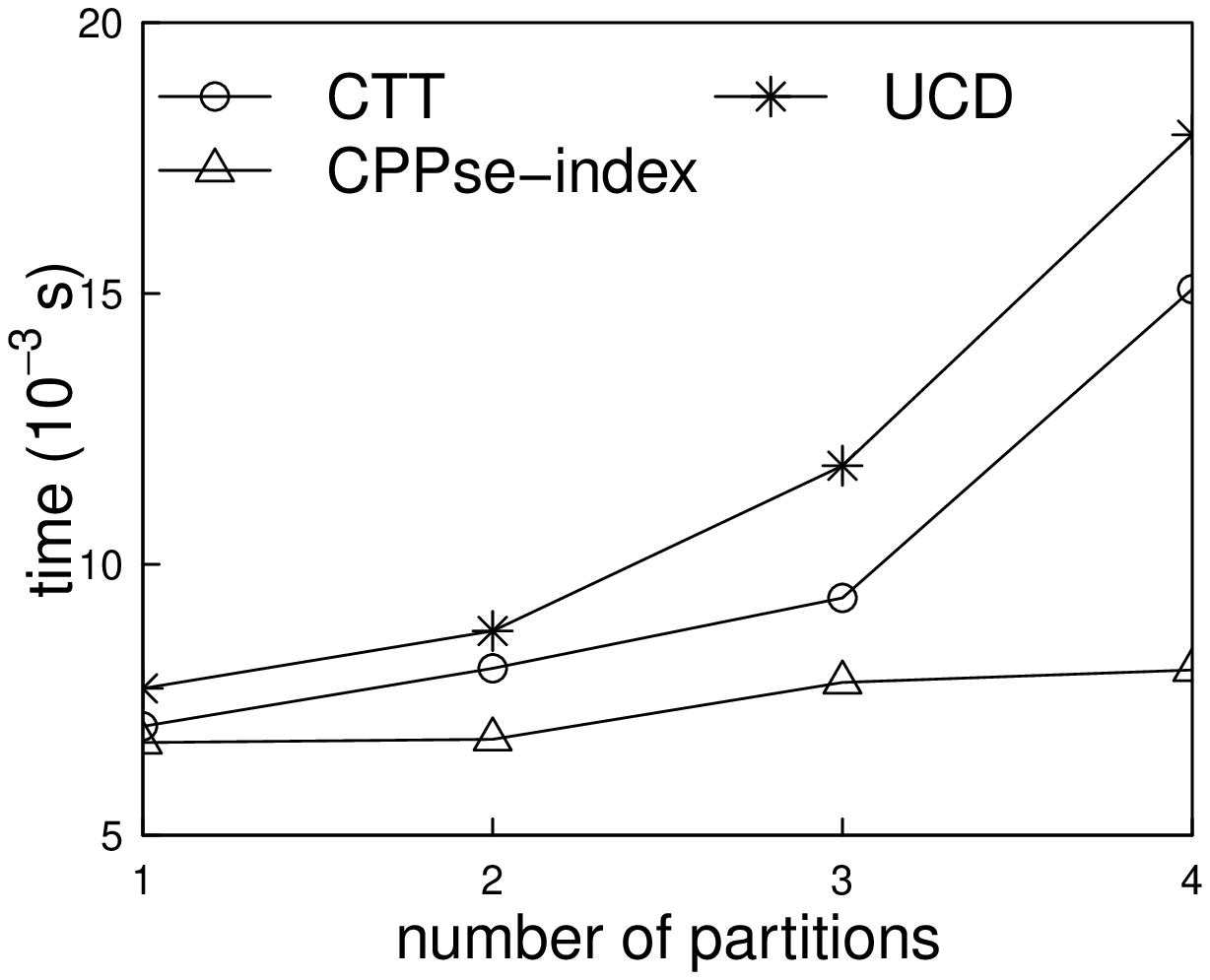}}
\subfigure[\syoutubec]{\label{fig:b}\includegraphics[width=0.46\linewidth]{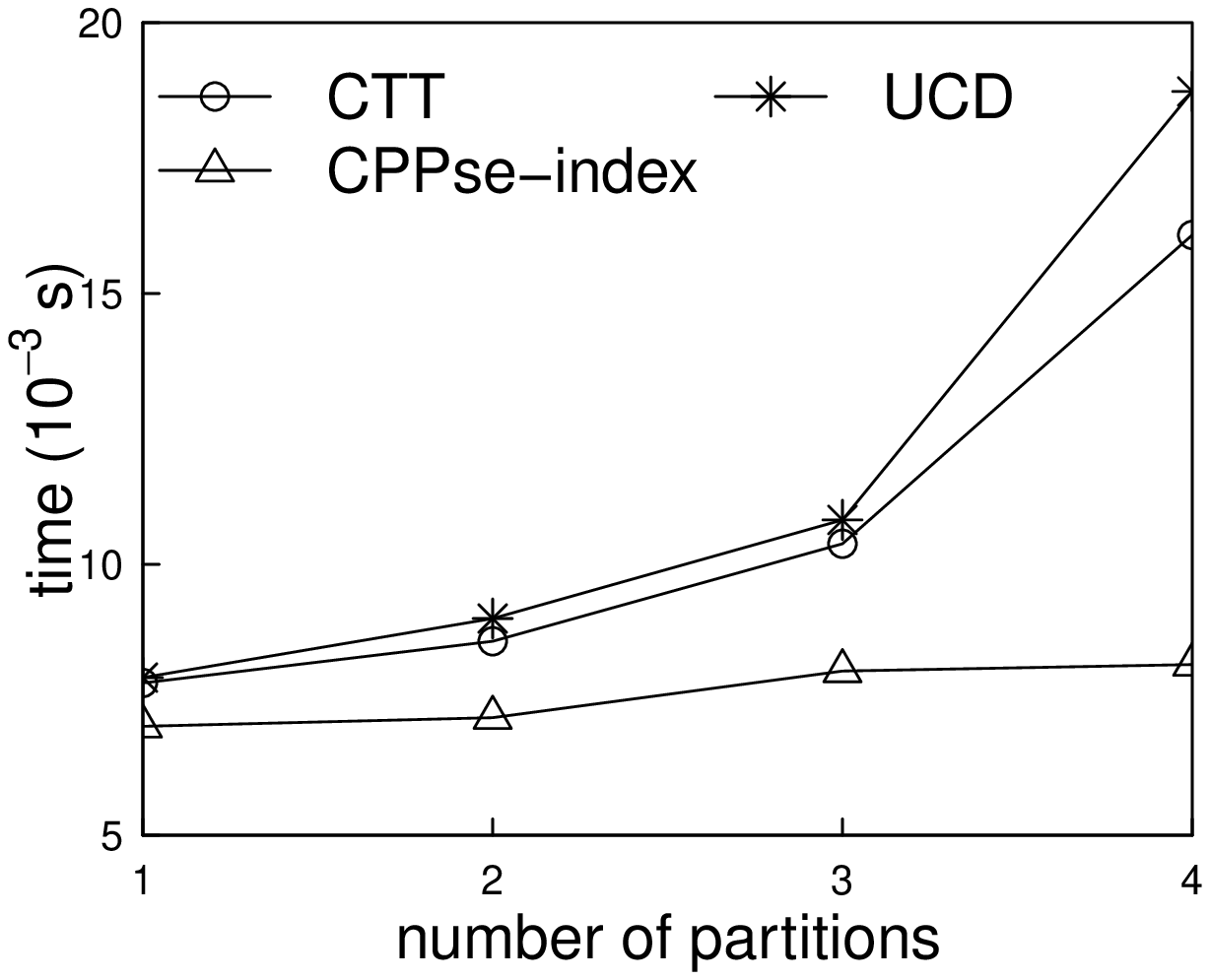}}
\subfigure[\movielenc]{\label{fig:b}\includegraphics[width=0.46\linewidth]{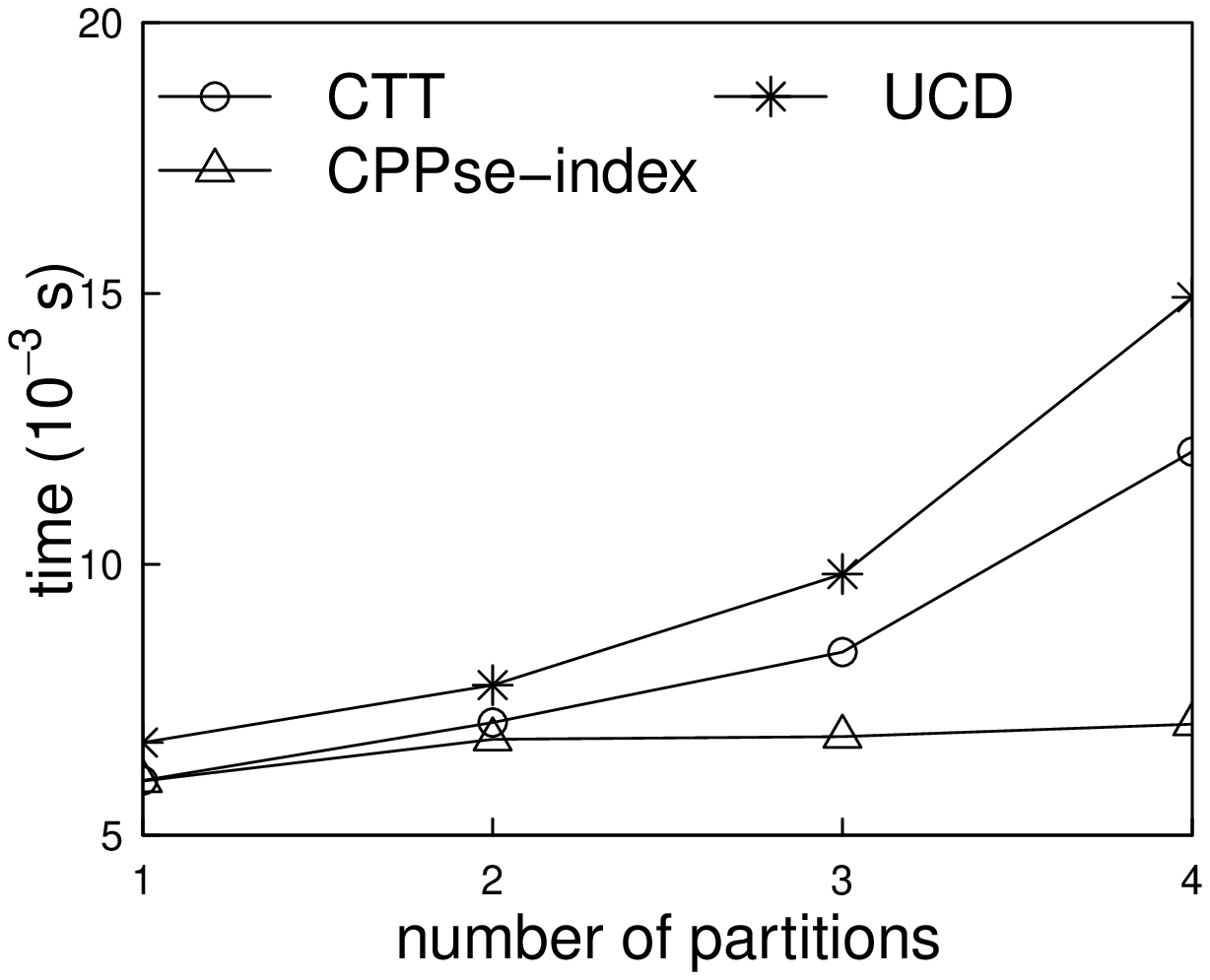}}
\subfigure[\smovielenc]{\label{fig:b}\includegraphics[width=0.46\linewidth]{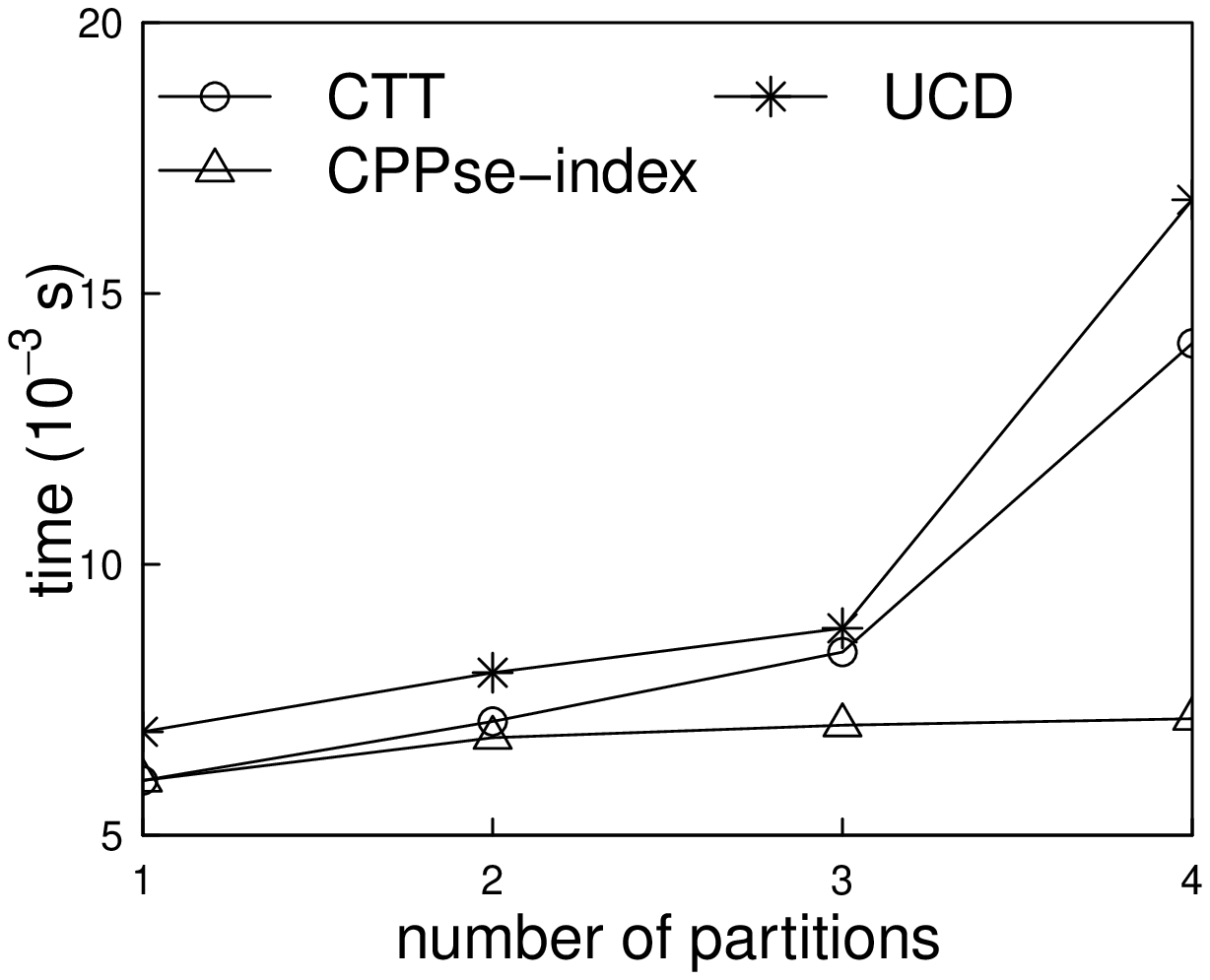}}\vspace{-1.5ex}
\caption{Efficiency comparison}
\label{fig:efficiency}
\end{figure}

\subsection{Efficiency Evaluation}
We evaluate the efficiency of our proposed {\Tree} in terms of its recommendation and update costs.
Our {\Tree} is implemented over Apache Storm, which is a real-time fault-tolerant distributed data processing system \cite{storm}.
The bolt in  Apache Storm is responsible for receiving inputs and works as the CPU. We configure the number of bolts over Apache Storm same as the category number of each dataset.

\subsubsection{Recommendation Efficiency Comparison}
We compare our proposed method with the state-of-the-art methods CTT and UCD in terms of the average response time per item on the stream. Here, $k$ is set to 30.
Fig.~\ref{fig:efficiency} shows the time cost of recommendation, where the number of partitions indicates the data set size in the simulation setting and the time cost is accumulated over the four test partitions.
Clearly, our approach is much faster than both CTT and UCD, especially when a large number of items are required to be recommended.
More importantly, the average recommendation cost of our proposed method is less affected by the size of items while the cost of both CTT and UCD increases almost exponentially to the item size.
This is because the {\Tree} representation prunes out the false alarm candidates in the user-item matching process, while the other two methods can only process all candidates sequentially. Moreover, UCD  performs worse than CTT due to the extra time cost from the diversity-based matching in it.

\subsubsection{Efficiency of Media Updates}
We test the cost of media updates over our CPPse-index by changing the size of updates. The time cost changes over different context updates are reported in Fig. \ref{fig:update}. Clearly, the cost increases steadily with the update size increase. This is because our CPPse-index processes the media updates with the support of hash scheme and user blocking techniques, which quickly locates the positions of the entries with user activity updates. This has proved our CPPse-index can be updated efficiently when the user profile updates happen.
\begin{figure}[h]
\centering
\subfigure[]{\label{fig:a}\includegraphics[width=0.46\linewidth]{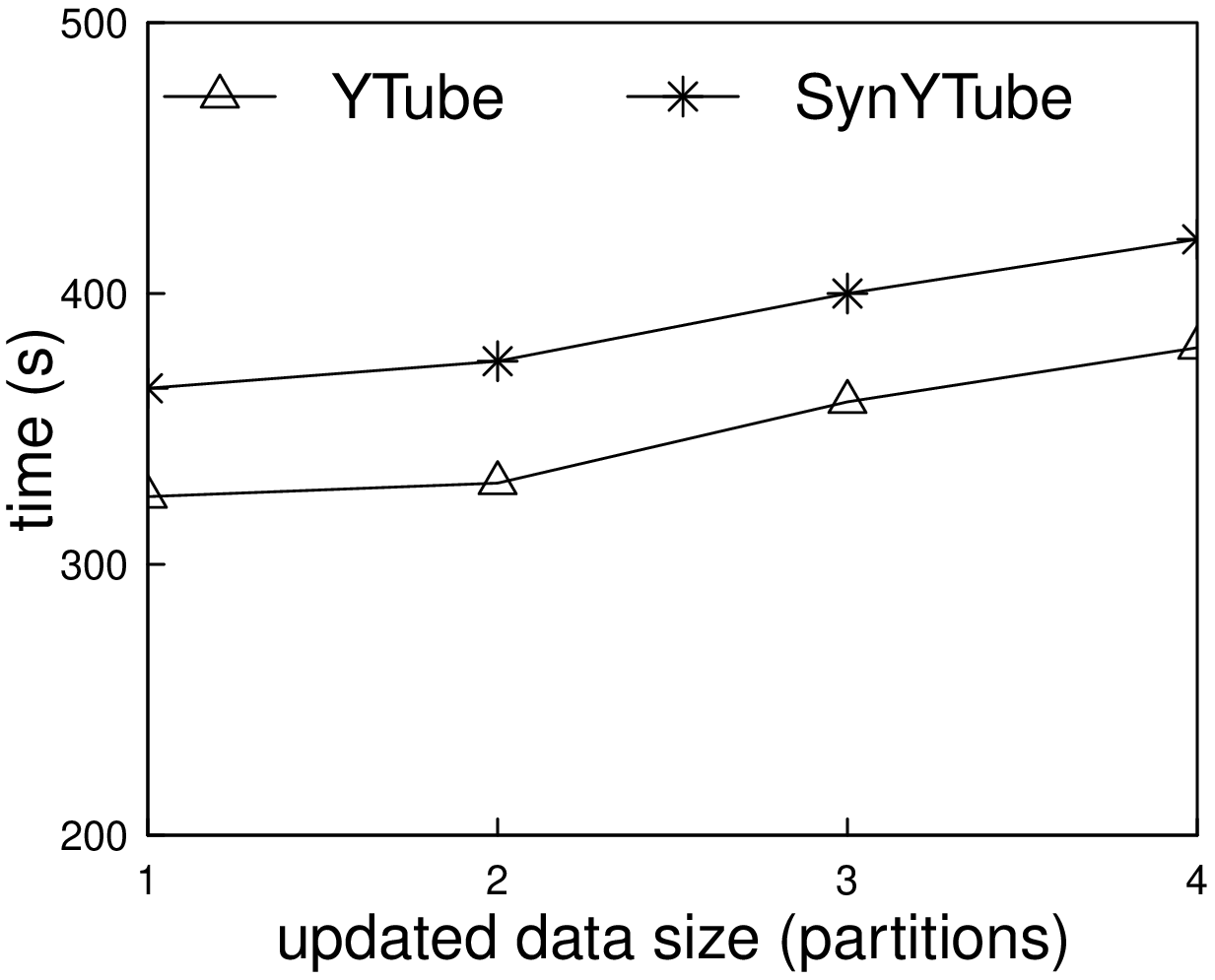}}
\subfigure[]{\label{fig:b}\includegraphics[width=0.46\linewidth]{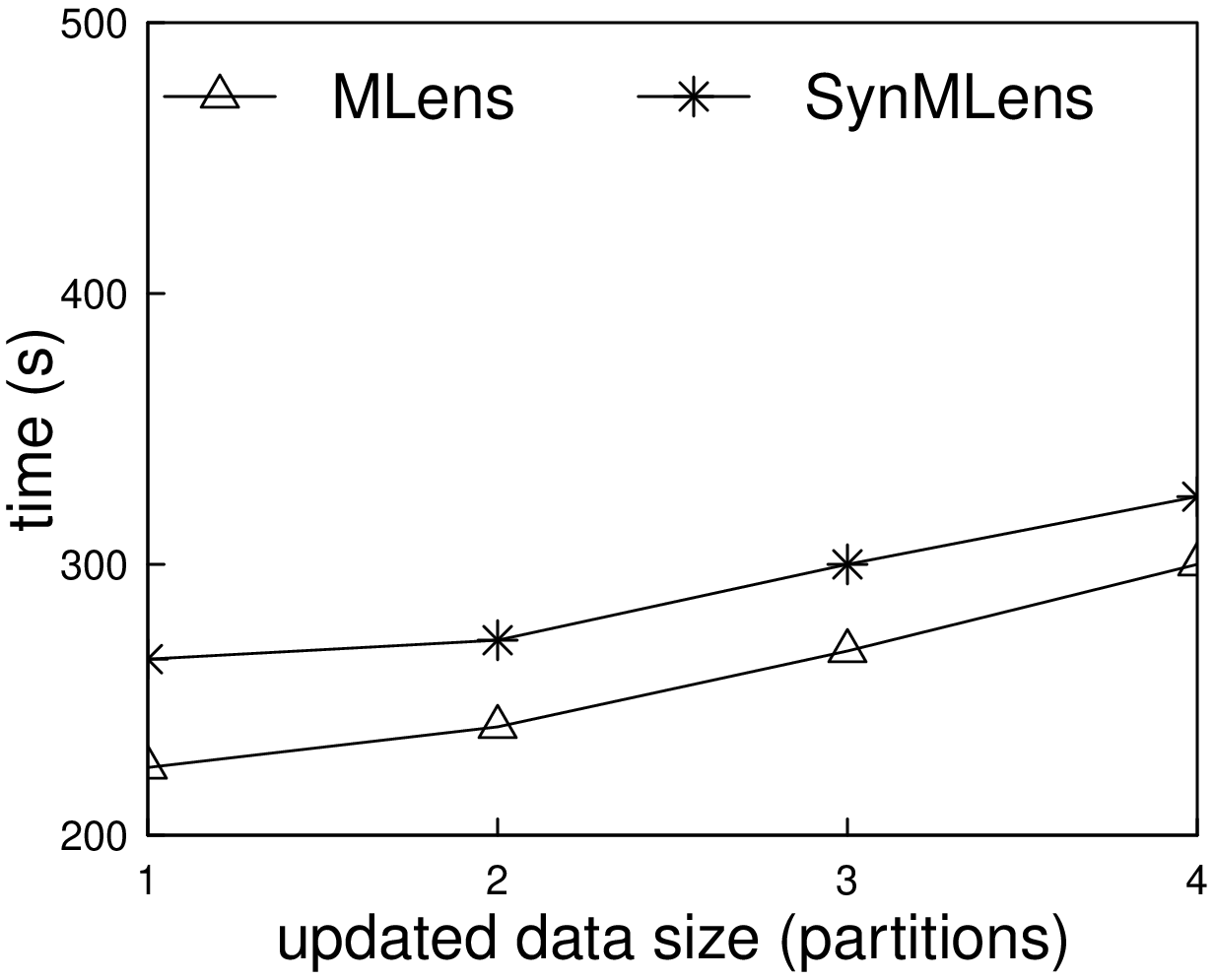}}\vspace{-2ex}
\caption{Efficency of social updates}\vspace{-0ex}
\label{fig:update}
\end{figure}

\section{CONCLUSION}\label{sec:conclusion}
This paper studies the problem of media stream recommendation. We first propose a novel Bi-Layer HMM model for predicting the users' long-term interest patterns. Then, we model both user profile and media data as streams, and propose a novel probability-based item-user matching approach. Finally, we propose an index scheme that optimizes the time cost of stream recommendation. The experimental results demonstrate the high effectiveness and efficiency of our proposed stream recommendation approach.

\begin{footnotesize}
	\setlength{\bibsep}{-0.1ex}
	\bibliographystyle{abbrvnat}

\end{footnotesize}

\end{document}